\documentclass[journal]{IEEEtran}
\usepackage{amsmath,amsfonts}
\usepackage{algorithmic}
\usepackage{algorithm}
\usepackage{array}
\usepackage[caption=false,font=normalsize,labelfont=sf,textfont=sf]{subfig}
\usepackage{textcomp}
\usepackage{stfloats}
\usepackage{url}
\usepackage[hidelinks]{hyperref}
\usepackage{orcidlink}
\usepackage{diagbox}  
\usepackage{verbatim}
\usepackage{graphicx}
\usepackage{bbm}
\usepackage{cite}
\begin{document}

\title{BeamGuard: Risk-Aware Multimodal Beam Forecasting and Adaptive Virtual Beamwidth Control for 6G mmWave V2I Links}


\author{IEEE Publication Technology,~\IEEEmembership{Staff,~IEEE,}
\author{Abidemi Orimogunje, 
Dejan Vukobratovic, Sunwoo Kim, Igbafe Orikumhi, Vukan Ninkovic, Evariste Twahirwa, and Gaspard Gashema,}

\thanks{This paper was produced by the IEEE Publication Technology Group. They are in Piscataway, NJ.}
\thanks{Manuscript received April 19, 2021; revised August 16, 2021.}}



\maketitle

\begin{abstract}
Reliable beam management is a central challenge for 6G millimeter-wave (mmWave) vehicle-to-infrastructure (V2I) links, where narrow beams provide high array gain but are vulnerable to mobility-induced misalignment, blockage, and domain variation. BeamGuard is a multimodal sensing-aided beam-management framework that combines exteroceptive sensing with optional partial in-band mmWave power observations to forecast future beam distributions and select adaptive virtual beamwidth actions for reliable V2I control. It fuses camera, radar, LiDAR, GPS, and mmWave power observations with a temporal multimodal forecaster, then converts the predicted posterior into a beam center and virtual codebook-level beamwidth through a risk-aware planner. Here, virtual beamwidth denotes adjacent-beam coverage in the codebook index space rather than physical analog wide-beam synthesis. BeamGuard supports sensor-only operation for beam-training overhead reduction, limited in-band operation with masked beam-power entries, and full hybrid operation with sensing and communication-side measurements. We evaluate BeamGuard on DeepSense 6G Scenarios 32 and 33, with additional held-out tests on Scenarios 31 and 34, covering day--night training, transfer, limited adaptation, ablations, budget sweeps, and lightweight baselines. The full-hybrid anchor, used as the complete-system reference, achieves Top-1/Top-3/Top-5 accuracies of approximately \(0.393/0.778/0.897\), while the planner attains a threshold-based outage probability of about \(0.0060\) with a gain ratio of about \(0.895\). Matched-budget baselines further show that BeamGuard improves over multilayer perceptron, recurrent, and temporal convolutional predictors under comparable in-band observation settings. These results demonstrate robust, overhead-aware beam management through multimodal forecasting and risk-aware virtual beamwidth control.
\end{abstract}

\section{Introduction}
\IEEEPARstart{M}{illimeter-wave} (mmWave) vehicle-to-infrastructure (V2I) communication is a key technology for high-throughput connected mobility, roadside sensing, cooperative perception, and low-latency vehicular services~\cite{b1}. By exploiting large bandwidth and highly directional antenna arrays, mmWave links can deliver substantial beamforming gain, but this gain depends critically on accurate and timely beam alignment \cite{b1,b2,b3}. In vehicular environments, rapid motion, blockage, and changing scattering geometry can cause the best narrow-beam index to vary quickly over time, while day--night sensing-domain shifts can reduce the reliability of the multimodal observations used for beam selection. Hence, reliable beam management requires more than selecting the strongest instantaneous beam; it requires forecasting future beam distributions and adapting beam coverage to the uncertainty of the prediction\cite{b4, b5, b6}. This motivates a beam-management framework that jointly learns future beam distributions from multimodal observations and converts those forecasts into beamwidth-aware decisions that control outage risk while preserving array gain.

Conventional mmWave beam management has largely relied on beam sweeping, hierarchical codebook search, beam tracking, and geometry-aided alignment \cite{b7, b8, b9, b10, b11}. These approaches are physically interpretable and compatible with analog or hybrid beamforming architectures, but they face a fundamental overhead-reliability tradeoff. Exhaustive search over a fine beam codebook improves alignment accuracy but incurs large training overhead, while narrow fixed beams provide high gain, but are fragile under mobility and blockage~\cite{b12}. Wider beams increase angular coverage and reduce misalignment probability, but they also sacrifice directional gain~\cite{b2}. This tradeoff is particularly severe in V2I links, where the roadside unit must maintain connectivity to fast-moving vehicles under limited beam-training time. Prior work on mmWave systems and hybrid beamforming has established the importance of directional beam alignment and low-overhead beam training~\cite{b13, b14, b15}, but robust beam selection under predictive uncertainty remains a challenging open problem.

Recent work has explored machine learning and deep learning for sensing-aided beam prediction using position, camera, radar, light detection and ranging (LiDAR), and multimodal observations to reduce beam-training overhead and improve beam selection in dynamic environments~\cite{b17,b18,b19,b20,b21,b22}. More recently, transformer-based and multimodal beam-prediction methods have extended this research direction to vision-position, LiDAR-aided, and missing-modality settings~\cite{b23a,b23b,b23c}, providing useful external context for the transformer-style forecasting component of BeamGuard. In parallel, large-scale multimodal resources such as DeepSense 6G have enabled real-world studies of sensing-aided communication using synchronized exteroceptive observations and mmWave measurements~\cite{b23}. These modalities provide complementary information: visual and geometric sensors capture environmental structure, GPS provides coarse position and mobility context, and in-band mmWave power measurements offer direct evidence of the beam response. However, most learning-based studies continue to emphasize top-\(K\) beam-prediction accuracy (top-\(K_{\mathrm{acc}}\)) as the primary objective. For practical communication control, the more important question is how a predicted future beam distribution should be converted into a beam action that balances link reliability, retained gain, switching cost, and beam-training overhead.

Bridging this gap between beam prediction and beam control requires addressing several unresolved limitations in existing sensing-aided beam-management approaches. First, many methods evaluate beam prediction under a single operating condition, making it difficult to assess robustness under day--night domain variation or limited target-domain supervision. Second, using the full mmWave power vector provides strong communication-side information, but obtaining it requires probing many or all beams in the codebook. Evaluating only this full-observation setting can therefore hide the beam-training overhead, making it necessary to also study partial in-band observations and sensor-only operation. Third, accuracy-only predictors do not directly address whether prediction uncertainty can be translated into reliable beam-control actions. Finally, limited modality ablations often obscure which sensing streams are responsible for performance gains. These limitations motivate a unified evaluation framework that compares sensor-only, in-band-only, and hybrid multimodal operation; studies partial mmWave power observations; and evaluates both prediction quality and communication-level control metrics such as outage, gain ratio, and switching rate.

This paper proposes \emph{BeamGuard}, a risk-aware multimodal beam forecasting and adaptive beamwidth-control framework for mmWave V2I links. BeamGuard uses a temporal multimodal forecaster to predict future beam distributions from camera, radar, LiDAR, GPS, and optional partial or full mmWave power observations. The predicted distribution is then used by an explicit planner that selects a beam center and virtual beamwidth according to the estimated outage risk, expected gain, and switching cost. The framework supports sensor-only operation for reduced beam-training overhead, in-band operation with limited beam measurements, and hybrid operation that combines sensing and communication-side evidence. The main contributions are summarized as follows:
\begin{itemize}
    \item We formulate mmWave V2I beam management as a multimodal forecasting-and-control problem in which future beam distributions are mapped to virtual beamwidth actions that jointly account for predicted beam-miscoverage risk, beamforming gain, and switching cost.

    \item We develop a unified BeamGuard framework which is mask-aware and supports sensor-only, in-band-only, and hybrid multimodal operation, including partial mmWave power observations through explicit beam-observation masks that preserve measurement-validity information.

    \item We provide a broad evaluation on DeepSense 6G Scenarios 32 and 33, with additional held-out scenario tests on Scenarios 31 and 34. The evaluation covers joint day--night training, in-domain and cross-scenario transfer, limited target-domain adaptation, focused modality ablations, beam-observation budget sweeps, sampling-policy sensitivity, fixed-width planner and same-forecaster controller ablations, risk-budget sweeps, and controlled lightweight predictor baselines.
\end{itemize}
The rest of this paper is organized as follows. Section~II presents the system model and formulates the risk-aware beam forecasting and beamwidth-control problem. Section~III describes the proposed BeamGuard framework, including the multimodal temporal forecaster, beam-observation masking strategy, calibration procedure, and risk-aware planner. Section~IV details the DeepSense 6G dataset preparation, evaluation protocols, ablation design, baselines, and performance metrics. Section~V presents and discusses the experimental results, including operating-regime comparison, beam-observation budget analysis, transfer behavior, risk-budget sensitivity, predictor baselines, and runtime feasibility. Finally, Section~VI concludes the paper and outlines future research directions.

\section{System Model and Problem Formulation}
\label{sec:sysmod}

This section formalizes the beam-management setting considered in this work. We consider a mmWave V2I link equipped with a finite receive-side beam codebook and assisted by synchronized multimodal observations from the surrounding environment~\cite{b23}. The objective is to use a short history of sensing and optional communication-side measurements to forecast future beam distributions and select a beamwidth-aware action that balances predicted beam-coverage risk, beamforming gain, beamwidth cost, and switching cost under a beam-observation budget.

\subsection{Codebook, Beam labels, and Multimodal Observations}

Consider a V2I mmWave link with a receive-side beam codebook
\begin{equation}
    \mathcal{F}=\{\mathbf{f}_1,\mathbf{f}_2,\ldots,\mathbf{f}_M\},
    \label{eq:codebook}
\end{equation}
where \(M\) is the number of codebook beams and \(\mathbf{f}_m\) denotes the \(m\)th narrow-beam codeword. At discrete time index \(t\), beam training over the codebook produces the measured receive-power vector
\begin{equation}
    \mathbf{p}_t =
    [p_{t,1},p_{t,2},\ldots,p_{t,M}]^{\top}
    \in\mathbb{R}_{+}^{M},
    \label{eq:measured_power_vector}
\end{equation}
where \(p_{t,m}\) is the measured received power obtained when beam \(\mathbf{f}_m\) is probed. The best narrow-beam label is
\begin{equation}
    m_t^\star = \arg\max_{m\in\{1,\ldots,M\}} p_{t,m}.
    \label{eq:best_beam}
\end{equation}
The label \(m_t^\star\) provides the supervised beam target, while the measured vector \(\mathbf{p}_t\) provides communication-side information about the codebook response. Future power quantities predicted by the model are denoted separately by \(\widehat{\mathbf{p}}_{t+\tau}\).

At the same time, the system observes exteroceptive sensing modalities
\begin{equation}
    \mathbf{x}_t =
    \big(
    \mathbf{x}^{\mathrm{C}}_t,
    \mathbf{x}^{\mathrm{R}}_t,
    \mathbf{x}^{\mathrm{Lid}}_t,
    \mathbf{x}^{\mathrm{G}}_t
    \big),
    \label{eq:multimodal_observation}
\end{equation}
where the superscripts denote camera, radar, LiDAR, and GPS observations, respectively. These modalities provide complementary information about scene geometry, mobility state, and the surrounding environment.

In hybrid and in-band operating modes, the controller may additionally observe a subset of the mmWave power vector. Let
\begin{equation}
    \mathbf{o}_t\in\{0,1\}^{M}, \qquad \|\mathbf{o}_t\|_0=L,
    \label{eq:observation_mask}
\end{equation}
be the binary beam-observation mask, where \(L\) is the number of observed beam-power entries. The mask \(\mathbf{o}_t\) specifies which beam-power entries are available to the forecaster. The system model imposes only the budget constraint \(\|\mathbf{o}_t\|_0=L\), while the specific mask-generation rule is defined in Sec.~\ref{sec:experi_sec}. The masked in-band observation is
\begin{equation}
    \widetilde{\mathbf{p}}_t = \mathbf{o}_t \odot \mathbf{p}_t,
    \label{eq:masked_power}
\end{equation}
where \(\odot\) denotes element-wise multiplication. The cases \(L=0\) and \(L=M\) correspond to no in-band power observation and full power-vector observation, respectively, while intermediate values represent reduced beam-observation budgets.

The predictor operates on a history window of length \(H\). At decision time \(t\), the available input history is
\begin{equation}
    \mathcal{I}_t =
    \left\{
    \left(\mathbf{x}_{i}, \widetilde{\mathbf{p}}_{i}, \mathbf{o}_{i}\right)
    : i=t-H+1,\ldots,t
    \right\},
    \label{eq:history}
\end{equation}
where \(i\) indexes samples in the history window and \(H\) denotes the number of past samples available to the forecaster. The operating regime determines which components of \(\mathcal{I}_t\) are active; for example, the power modality is inactive when \(L=0\). 

\subsection{Beam Forecasting, Virtual Coverage, and Beam-miscoverage risk}

Given \(\mathcal{I}_t\), the forecaster predicts future beam posteriors over a horizon of length \(T\). For each future step \(\tau=1,\ldots,T\), the model outputs
\begin{equation}
    \widehat{\mathbf{q}}_{t+\tau}
    =
    [\widehat{q}_{t+\tau,1},\ldots,\widehat{q}_{t+\tau,M}]^{\top},
    \qquad
    \sum_{m=1}^{M}\widehat{q}_{t+\tau,m}=1 .
    \label{eq:beam_posterior}
\end{equation}
Here, \(\widehat{q}_{t+\tau,m}\ge 0\) is the predicted probability that beam \(m\) is the best beam at future time \(t+\tau\). 
The model also predicts an auxiliary power vector
\begin{equation}
    \widehat{\mathbf{p}}_{t+\tau}\in\mathbb{R}_{+}^{M},
    \label{eq:power_forecast}
\end{equation}
which encourages communication-consistent representations and captures relative beam-power structure beyond the best-beam label.

The planner selects a center beam \(c_t\in\{1,\ldots,M\}\) and a virtual beamwidth \(w_t\in\mathcal{W}\), where \(\mathcal{W}\subset\mathbb{N}_{+}\) is a finite set of allowable positive integer beamwidths. Odd beamwidths are used so that the covered codebook region can be centered symmetrically around \(c_t\). In this work, virtual beamwidth refers to the number of adjacent codebook beams covered by the control action, not to a specific analog wide-beam synthesis procedure.

For a selected action \((c_t,w_t)\), the covered codebook set is
\begin{equation}
    \mathcal{V}(c_t,w_t)
    =
    \left\{
    m\in\{1,\ldots,M\}:
    d_{\mathrm{idx}}(m,c_t)
    \leq
    \left\lfloor\frac{w_t-1}{2}\right\rfloor
    \right\},
    \label{eq:virtual_set}
\end{equation}
where
\begin{equation}
    d_{\mathrm{idx}}(m,c)=|m-c|
    \label{eq:index_distance}
\end{equation}
is the beam-index distance on the ordered sector codebook. This non-circular distance is used because the DeepSense mmWave receiver employs a 64-beam codebook that scans a finite field of view rather than a cyclic angular domain. Hence, edge beams are not wrapped, and virtual coverage near codebook boundaries is clipped to the available index set \(\{1,\ldots,M\}\).

Although the forecaster predicts \(T\) future steps, the receding-horizon planner executes the first action using the next-step posterior. The predicted probability that action \((c,w)\) covers the next-step best beam is
\begin{equation}
    P_t^{\mathrm{cov}}(c,w)
    =
    \sum_{m\in\mathcal{V}(c,w)}
    \widehat{q}_{t+1,m}.
    \label{eq:coverage_prob}
\end{equation}
The corresponding predicted beam-miscoverage risk is
\begin{equation}
    \rho_t(c,w)
    =
    1-P_t^{\mathrm{cov}}(c,w).
    \label{eq:miscoverage_risk}
\end{equation}
The quantity \(\rho_t(c,w)\) is used as the planner's risk proxy. Thus, actions that cover more posterior mass have lower predicted miscoverage risk, while narrow actions may preserve gain but are more sensitive to forecast uncertainty. The threshold-based empirical link outage metric used for final evaluation is defined separately in Sec.~\ref{sec:experi_sec}.

\subsection{Risk-aware Beamwidth control problem}

BeamGuard selects one beamwidth-aware action at each decision time using the forecasted posterior and auxiliary power prediction. To reduce search complexity while retaining strong candidate actions, the planner forms a finite candidate-center set \(\mathcal{C}_t\) from the highest-ranking entries of the next-step beam posterior and the predicted auxiliary power vector:
\begin{equation}
    \mathcal{C}_t =
    \mathrm{uniq}
    \left(
    \mathrm{Top}_{K_c}(\widehat{\mathbf{q}}_{t+1})
    \cup
    \mathrm{Top}_{K_c}(\widehat{\mathbf{p}}_{t+1})
    \right),
    \label{eq:candidate_centers}
\end{equation}
where \(K_c\) is the number of candidate center beams retained for planning, \(\mathrm{Top}_{K_c}(\cdot)\) returns the indices of the \(K_c\) largest entries, and \(\mathrm{uniq}(\cdot)\) removes duplicate centers. 
The candidate action set is
\begin{equation}
    \mathcal{A}_t =
    \left\{(c,w): c\in\mathcal{C}_t,\; w\in\mathcal{W}\right\}.
    \label{eq:action_set}
\end{equation}

The control objective is to choose an action that provides high predicted coverage and gain while discouraging excessive beamwidth, frequent switching, and high predicted miscoverage risk. For an action \((c,w)\in\mathcal{A}_t\), BeamGuard uses the score
\begin{equation}
\begin{aligned}
    J_t(c,w)
    &=
    \eta(w)\widehat{g}_t(c,w)
    -\lambda_w(w-1)
    -\lambda_{\rho}\left[\rho_t(c,w)-\beta\right]_+ \\
    &\quad
    -\lambda_{\mathrm{sw}}\mathbf{1}\{c\neq c_{t-1}\},
\end{aligned}
\label{eq:planner_score}
\end{equation}
where
\begin{equation}
    \widehat{g}_t(c,w)
    =
    \max_{m\in\mathcal{V}(c,w)}
    \widehat{p}_{t+1,m}.
    \label{eq:predicted_action_gain}
\end{equation}
Here, \(\widehat{g}_t(c,w)\) is the predicted action gain obtained from the auxiliary power forecast, \(\eta(w)\) is a width-dependent gain factor, and \([x]_+=\max(0,x)\). The term \(\lambda_w(w-1)\) penalizes wider virtual beams, \(\lambda_{\mathrm{sw}}\) penalizes center switching, and \(\lambda_{\rho}[\rho_t(c,w)-\beta]_+\) penalizes only the amount by which the predicted miscoverage risk exceeds the configured risk budget \(\beta\). The exact model, loss, calibration, and planner hyperparameters used in the experiments are specified in Table~\ref{tab:model_planner_settings}.
The conservative feasible set is
\begin{equation}
    \mathcal{A}_t^{\rho}
    =
    \left\{
    (c,w)\in\mathcal{A}_t:
    \rho_t(c,w)\leq \rho_{\max}
    \right\},
    \label{eq:risk_feasible_set}
\end{equation}
where \(\rho_{\max}\) is the hard risk-screening threshold. The selected action is
\begin{equation}
(c_t,w_t)=
\begin{cases}
\displaystyle
\arg\max_{(c,w)\in\mathcal{A}_t^{\rho}} J_t(c,w),
& \mathcal{A}_t^{\rho}\neq\emptyset, \\[1.2ex]
\displaystyle
\arg\min_{(c,w)\in\mathcal{A}_t} \rho_t(c,w),
& \mathcal{A}_t^{\rho}=\emptyset .
\end{cases}
\label{eq:planner_decision}
\end{equation}
The fallback rule prevents undefined actions while preserving risk-minimizing behavior when no candidate satisfies the conservative screen.

The resulting problem is a receding-horizon beam-control problem. The predictor forecasts future beam distributions over \(T\) steps, while the planner executes the first action based on the next-step posterior and repeats the process at the following time instant. This formulation connects multimodal beam prediction directly to communication control by evaluating not only beam-classification accuracy, but also predicted beam coverage, retained gain, beamwidth, and switching behavior.

\section{BEAMGUARD: Multimodal Forecasting and Risk-aware Beamwidth control}
\label{sec:beamguard_method}

\begin{figure*}[t]
\centering
\includegraphics[width=0.98\textwidth]{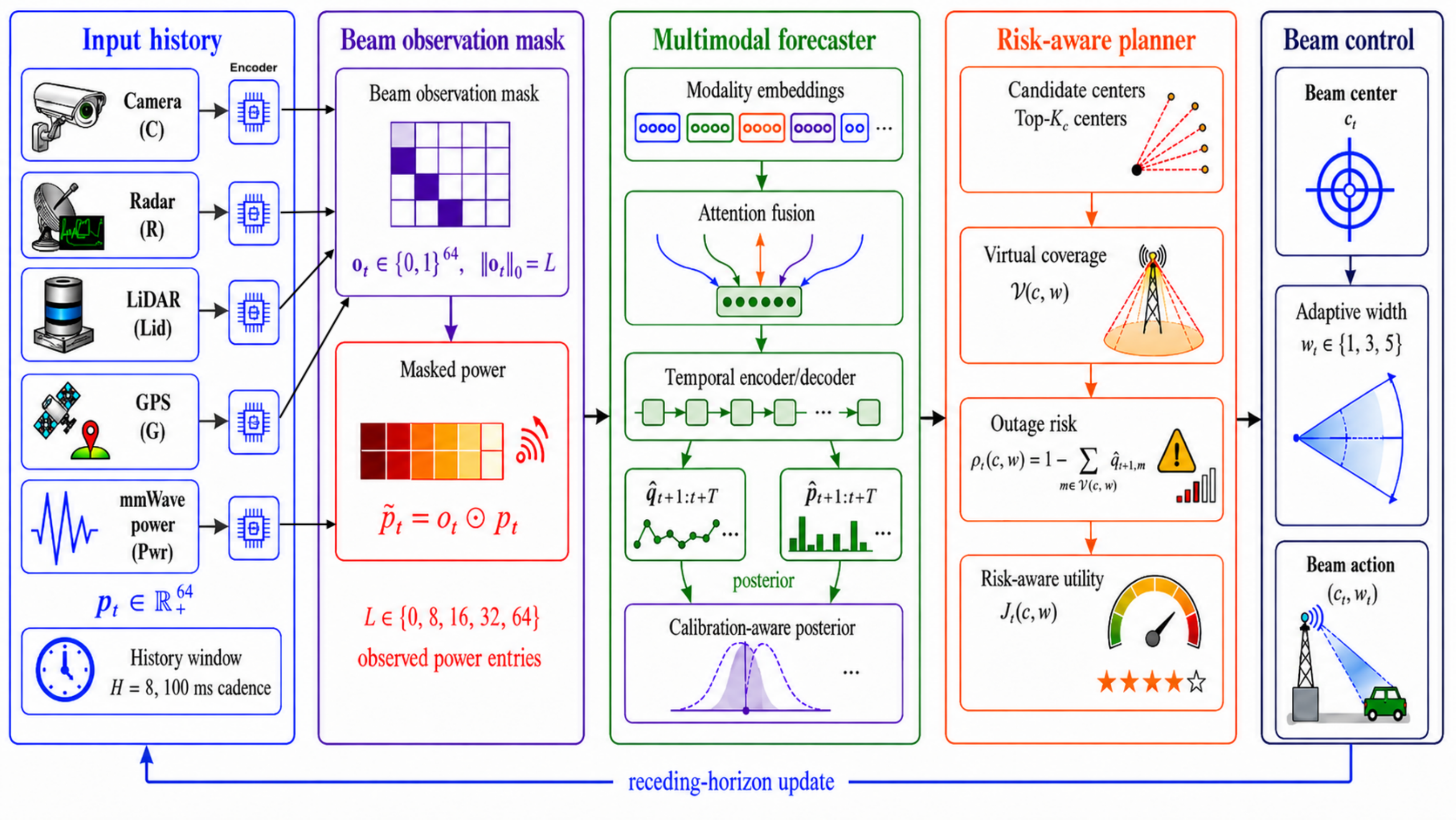}
\vspace{-2mm}
\caption{BeamGuard system overview. A history of synchronized camera, radar, LiDAR, GPS, and optional observed mmWave beam-power measurements is processed by modality-specific encoders and a temporal multimodal forecaster. The beam-observation mask $\mathbf{o}_t$ selects $L$ entries from the $M$-beam power vector, enabling sensor-only operation, partial in-band observation, and full hybrid operation within the same framework. The predicted beam posterior and auxiliary power forecast are used by a risk-aware planner to evaluate candidate beam centers, virtual beamwidths, posterior coverage, beam-miscoverage risk, predicted gain, and switching cost. The final control action consists of a selected beam center $c_t$ and adaptive virtual beamwidth $w_t$. Here, C, R, Lid, G, and Pwr denote camera, radar, LiDAR, GPS, and mmWave beam power, respectively, while $L$ denotes the beam-observation budget.}
\label{fig:system_overview}
\end{figure*}

BeamGuard connects sensing-aided beam forecasting with communication-aware beam control. Rather than treating beam selection as a standalone classification task, the framework first estimates future beam distributions from the history in \eqref{eq:history}, then uses the selected posterior to choose a beam center and virtual beamwidth according to the risk-aware score in \eqref{eq:planner_score}. This design addresses the overhead--reliability tradeoff in mmWave V2I links: sensor-only operation avoids in-band beam-power measurements, partial in-band operation limits the number of measured beams, and hybrid operation combines exteroceptive sensing with communication-side evidence for reliability-oriented control. Fig.~\ref{fig:system_overview} illustrates the complete BeamGuard processing flow.

\vspace{-2mm}
\subsection{Multimodal Temporal Forecaster}

The forecasting module maps the observation history $\mathcal{I}_t$ in \eqref{eq:history} to future beam posteriors $\{\widehat{\mathbf{q}}_{t+\tau}\}_{\tau=1}^{T}$ and auxiliary power predictions $\{\widehat{\mathbf{p}}_{t+\tau}\}_{\tau=1}^{T}$. Let $\mathcal{U}_{\mathrm{x}}=\{\mathrm{C},\mathrm{R},\mathrm{Lid},\mathrm{G}\}$ denote the exteroceptive sensing modalities, and let $\mathcal{U}=\mathcal{U}_{\mathrm{x}}\cup\{\mathrm{pwr}\}$ denote the full modality set, where $\mathrm{pwr}$ represents the mmWave beam-power modality. For a given operating regime, $\mathcal{U}_{\mathrm{act}}\subseteq\mathcal{U}$ denotes the active modality set used by the forecaster.

At each history index $i$, each active modality is mapped to a common latent dimension. For modality $u$, the encoder input is denoted by $\mathbf{s}_i^u$. For sensing modalities, $\mathbf{s}_i^u=\mathbf{x}_i^u$ for $u\in\mathcal{U}_{\mathrm{x}}$. For the mmWave power modality, the encoder input is the concatenation of the masked power vector and the binary observation mask,
\begin{equation}
    \mathbf{s}_{i}^{\mathrm{pwr}}
    =
    [\widetilde{\mathbf{p}}_{i};\mathbf{o}_{i}],
    \label{eq:power_encoder_input}
\end{equation}
where $[\cdot;\cdot]$ denotes concatenation. This representation preserves both the measured beam-power values and their measurement-validity indicators. Each active modality is encoded as
\begin{equation}
    \mathbf{h}_{i}^{u}=E_u(\mathbf{s}_{i}^{u})\in\mathbb{R}^{d},
    \qquad u\in\mathcal{U}_{\mathrm{act}},
    \label{eq:modality_embedding}
\end{equation}
where $E_u(\cdot)$ is the encoder for modality $u$, and $d$ is the common embedding dimension.
The modality embeddings are fused into one temporal token,
\begin{equation}
    \mathbf{z}_i
    =
    F_{\mathrm{fuse}}
    \left(
    \{\mathbf{h}_{i}^{u}:u\in\mathcal{U}_{\mathrm{act}}\}
    \right)
    \in\mathbb{R}^{d},
    \label{eq:fusion_token}
\end{equation}
where $F_{\mathrm{fuse}}(\cdot)$ denotes the multimodal fusion block. The resulting history-token sequence is
\begin{equation}
    \mathbf{Z}^{\mathrm{hist}}_t
    =
    [\mathbf{z}_{t-H+1},\ldots,\mathbf{z}_{t}]^{\top}
    \in\mathbb{R}^{H\times d}.
    \label{eq:token_sequence}
\end{equation}
This sequence is processed by the temporal encoder and decoded over the forecasting horizon. Thus, $\mathbf{x}_i^u$, $\mathbf{s}_i^u$, $\mathbf{h}_i^u$, and $\mathbf{z}_i$ denote the sensing observation, encoder input, modality embedding, and fused temporal token respectively.

An autoregressive decoding head maps the temporal representation to beam logits $\boldsymbol{\ell}_{t+\tau}\in\mathbb{R}^{M}$ and auxiliary power predictions $\widehat{\mathbf{p}}_{t+\tau}\in\mathbb{R}_{+}^{M}$ for $\tau=1,\ldots,T$. The beam posterior is obtained as
\begin{equation}
    \widehat{\mathbf{q}}_{t+\tau}
    =
    \mathrm{softmax}(\boldsymbol{\ell}_{t+\tau}).
    \label{eq:decoder_posterior}
\end{equation}
The posterior $\widehat{\mathbf{q}}_{t+\tau}$ is used by the planner, while the auxiliary power prediction encourages the latent representation to preserve the relative beam-power structure of the mmWave codebook. This is important because the best-beam label in \eqref{eq:best_beam} contains only the index of the strongest beam, whereas the full power response carries information about neighboring beams and angular uncertainty.
\vspace{-2mm}
\subsection{Beam-observation masking and Operating Modes}

BeamGuard realizes different operating modes by changing the active modality set while preserving the same forecasting pipeline. In the sensor-only mode, the power modality is inactive and the forecaster relies on camera, radar, LiDAR, and GPS observations. In the in-band-only mode, the forecaster uses the mmWave power modality without exteroceptive sensing. In the hybrid mode, the model fuses exteroceptive sensing with partial or full in-band power observations. For partial in-band observation $(0<L<M)$, the power encoder receives both $\widetilde{\mathbf{p}}_t$ and $\mathbf{o}_t$. The mask encodes measurement validity rather than signal strength, preventing unobserved codebook entries from being treated as measured low-power beams.

This masking mechanism is part of the model representation, not a preprocessing detail. The power input retains the $M$-dimensional codebook structure, while the observation mask indicates which entries are valid measurements under the current beam-observation budget. As a result, the same architecture can operate in sensor-only, partial in-band, and full hybrid settings, enabling BeamGuard to trade beam-training overhead against communication reliability without changing the forecasting model.
\vspace{-2mm}
\subsection{TRAINING OBJECTIVE}
The forecaster is trained using a multi-task objective over the $T$-step prediction horizon. For each training window, the beam-classification loss is
\begin{equation}
    \mathcal{L}_{\mathrm{ce}}
    =
    -\frac{1}{T}
    \sum_{\tau=1}^{T}
    \log
    \widehat{q}_{t+\tau,m_{t+\tau}^{\star}},
    \label{eq:ce_loss}
\end{equation}
where $m_{t+\tau}^{\star}$ is the measured best-beam label at future step $t+\tau$. The auxiliary power loss is
\begin{equation}
    \mathcal{L}_{\mathrm{pow}}
    =
    \frac{1}{TM}
    \sum_{\tau=1}^{T}
    \left\|
    \widehat{\mathbf{p}}_{t+\tau}
    -
    \mathbf{p}_{t+\tau}
    \right\|_{1},
    \label{eq:power_loss}
\end{equation}
which encourages the latent representation to preserve the relative beam-power structure of the codebook.

A beam-distance-aware regularizer discourages probability mass on beams that are far from the measured best beam. We define the distance weight
\begin{equation}
    \omega_{\sigma}(m,m^\star)
    =
    1-\exp\!\left(
    -\frac{d_{\mathrm{idx}}(m,m^\star)^2}{2\sigma^2}
    \right),
    \label{eq:distance_weight}
\end{equation}
where \(d_{\mathrm{idx}}(m,m^\star)=|m-m^\star|\) is the ordered beam-index distance from \eqref{eq:index_distance}, and \(\sigma>0\) controls how rapidly the penalty increases with beam-index separation. The distance-aware loss is then
\begin{equation}
    \mathcal{L}_{\mathrm{dist}}
    =
    \frac{1}{T}
    \sum_{\tau=1}^{T}
    \sum_{m=1}^{M}
    \omega_{\sigma}(m,m_{t+\tau}^{\star})
    \widehat{q}_{t+\tau,m}.
    \label{eq:distance_loss}
\end{equation}
The total training objective is
\begin{equation}
    \mathcal{L}_{\mathrm{tr}}
    =
    \mathcal{L}_{\mathrm{ce}}
    +
    \alpha_{p}\mathcal{L}_{\mathrm{pow}}
    +
    \alpha_{d}\mathcal{L}_{\mathrm{dist}},
    \label{eq:training_loss}
\end{equation}
where $\alpha_p$ and $\alpha_d$ control the auxiliary power and distance-aware terms. Model selection is performed on the validation split, and final performance is reported only on the held-out test split.

\subsection{Calibration and Posterior selection}

Due to the planner's reliance on posterior mass for estimating coverage and beam-miscoverage risk, calibration is applied as a post-processing step. A scalar temperature $\zeta>0$ is fitted on the calibration split by minimizing negative log-likelihood (NLL):
\begin{equation}
    \zeta^{\star}
    =
    \arg\min_{\zeta>0}
    \frac{1}{N_{\mathrm{cal}}}
    \sum_{n=1}^{N_{\mathrm{cal}}}
    -
    \log
    \left[
    \mathrm{softmax}
    \left(
    \frac{\boldsymbol{\ell}_{n}}{\zeta}
    \right)
    \right]_{m_n^\star},
    \label{eq:temperature_fit}
\end{equation}
where $N_{\mathrm{cal}}$ is the number of calibration samples after flattening over prediction windows and future steps, $\boldsymbol{\ell}_n$ is the beam-logit vector, and $m_n^\star$ is the corresponding measured best-beam label. The calibrated posterior is
\begin{equation}
    \widehat{\mathbf{q}}^{\,\mathrm{cal}}_n
    =
    \mathrm{softmax}
    \left(
    \frac{\boldsymbol{\ell}_n}{\zeta^{\star}}
    \right).
    \label{eq:temperature_scaling}
\end{equation}

After fitting $\zeta^{\star}$, BeamGuard compares the raw and calibrated posteriors on the validation split using expected calibration error (ECE) and NLL. The calibrated posterior is used when it improves validation ECE without increasing validation NLL; otherwise, the raw posterior is retained. This selection is performed before test evaluation and prevents post-hoc calibration from degrading models whose raw posteriors are already well behaved.
\vspace{-2mm}
\subsection{RISK-AWARE BEAMWIDTH PLANNER}

The planner converts the next-step posterior and auxiliary power forecast into a communication action. Candidate center beams are obtained from the unique union of the top $K_c$ entries of the next-step beam posterior and the top $K_c$ entries of the predicted auxiliary power vector, as defined in \eqref{eq:candidate_centers}. The planner then evaluates each retained center with each allowable virtual beamwidth using the covered set in \eqref{eq:virtual_set}. The predicted coverage probability $P_t^{\mathrm{cov}}(c,w)$, beam-miscoverage risk $\rho_t(c,w)$, and predicted action gain $\widehat{g}_t(c,w)$ are computed from \eqref{eq:coverage_prob}, \eqref{eq:miscoverage_risk}, and \eqref{eq:predicted_action_gain}, respectively.

The planner solves \eqref{eq:planner_decision} by deterministic enumeration over the finite candidate action set $\mathcal{A}_t$. Since candidate centers are restricted to the unique union of posterior-ranked and predicted-power-ranked beams, the planner evaluates at most $|\mathcal{C}_t||\mathcal{W}|\leq 2K_c|\mathcal{W}|$ center--width actions per decision step before duplicate centers are removed. With $K_c=10$ and $\mathcal{W}=\{1,3,5\}$, this corresponds to at most $60$ candidate actions, and often fewer after applying $\mathrm{uniq}(\cdot)$ in \eqref{eq:candidate_centers}. With direct summation over the covered set $\mathcal{V}(c,w)$, the per-step computational cost is
\begin{equation}
    \mathcal{O}\!\left(
    |\mathcal{C}_t|\sum_{w\in\mathcal{W}}w
    \right),
    \label{eq:planner_complexity}
\end{equation}
which is negligible compared with multimodal feature extraction and temporal forecasting.

This planner differs from greedy narrow-beam selection in two ways. First, it does not assume that the most likely beam index is always the best communication action. If the predicted posterior is spread across adjacent beams, a wider virtual beamwidth can reduce miscoverage risk even when it slightly reduces gain. Second, the planner explicitly penalizes excessive switching and overly wide coverage. The resulting action balances reliability and efficiency; narrow beams are preferred when the forecast is confident, while wider coverage may be selected when predicted uncertainty is high.

At each time step, BeamGuard executes a receding-horizon procedure. First, synchronized sensing and optional in-band measurements over the history window are encoded and fused. Second, the temporal forecaster produces future beam posteriors and auxiliary power predictions. Third, the posterior selected by the calibration rule is passed to the planner. Finally, the planner evaluates candidate center--width pairs and executes the first action. The same process is repeated at the next time instant using the updated observation history.

This modular design separates forecasting from control. The forecaster can be evaluated using top-$K_{\mathrm{acc}}$ accuracy, calibration, beam-distance agreement, and auxiliary power-regression metrics, while the planner can be evaluated using communication-level metrics such as threshold-based outage probability, gain ratio, and switching rate. The same modularity enables the focused ablations used in the experiments, including sensor-only operation, partial mmWave beam budgets, multimodal fusion, and risk-budget sensitivity.
\vspace{-2mm}
\section{Experimental setup, Ablations and Evaluation Metrics}
\label{sec:experi_sec}

The evaluation tests whether BeamGuard forecasts future beam distributions accurately and converts those forecasts into reliable beam-control actions. The experiments cover joint day--night training, in-domain and cross-scenario transfer, limited target-domain adaptation, beam-observation budgets, operating regimes, risk-budget sensitivity, and lightweight predictor baselines.
\vspace{-2mm}
\subsection{Dataset and Preprocessing}

The experiments use DeepSense 6G Scenarios 32 and 33, which provide synchronized V2I measurements from camera, radar, LiDAR, GPS, and mmWave beam training~\cite{b24,b25}. Scenario 32 corresponds to daytime V2I data collection, while Scenario 33 captures the same environment under nighttime conditions. This paired structure supports evaluation under both in-domain testing and day--night domain transfer.

Each synchronized frame is associated with a measured receive-power vector \(\mathbf{p}_t\in\mathbb{R}_{+}^{M}\) over the \(M=64\) beam codebook defined in \eqref{eq:codebook} and \eqref{eq:measured_power_vector}. The best-beam label \(m_t^\star\) is obtained from \eqref{eq:best_beam}. The raw sensor streams are standardized into a common manifest containing frame indices, scenario identifiers, segment identifiers, modality paths, best-beam labels, and power-vector paths.

All data splits are constructed at the segment level rather than by random frame sampling. This prevents adjacent frames from the same continuous trajectory from appearing in both training and test partitions. The standardized dataset contains 6600 synchronized rows, with 3000 rows from Scenario 32 and 3600 rows from Scenario 33. Under the joint day--night protocol, the split contains 4000 training rows, 800 validation rows, 800 calibration rows, and 1000 test rows. After applying the history and horizon windowing, this corresponds to 3760 training windows, 752 validation windows, 752 calibration windows, and 940 test windows.

Separate validation, calibration, and test partitions are used. The validation partition selects model checkpoints, hyperparameters, and the posterior mode used by the planner. The calibration partition fits the temperature-scaling parameter, and the held-out test partition is used only for final reporting. This separation is important because the planner relies on posterior mass in \eqref{eq:coverage_prob} to estimate coverage and beam-miscoverage risk. The preprocessing and normalization settings used to construct the multimodal training windows are summarized in Table~\ref{tab:preprocess_baseline_summary}.

\begin{table*}[!t]
\caption{Input Preprocessing and Predictor Architecture Summary}
\label{tab:preprocess_baseline_summary}
\centering
\scriptsize
\setlength{\tabcolsep}{2.5pt}
\renewcommand{\arraystretch}{1.04}
\begin{tabular}{|p{92pt}|p{397pt}|}
\hline
\multicolumn{2}{|l|}{\textbf{Input Preprocessing and Normalization}} \\
\hline
\textbf{Input} & \textbf{Preprocessing / Normalization} \\
\hline
Camera &
RGB frames are resized to \(224\times224\), converted to
\(\mathrm{float32}\), and scaled by \(1/255\). \\
\hline
Radar &
Radar maps are converted to \(\mathrm{float32}\), with NaN and Inf
entries replaced by zero; each map is resized to \(128\times128\),
map-normalized, and represented using \(4\) channels. \\
\hline
LiDAR &
LiDAR maps are converted to \(\mathrm{float32}\), with NaN and Inf
entries replaced by zero; each map is resized to \(128\times128\),
map-normalized, and represented using \(1\) channel. \\
\hline
GPS &
Each GPS observation is represented by a \(4\)-D vector. NaN and
Inf entries are replaced by zero, and the four components are scaled
by \([50,50,10,10]\). \\
\hline
mmWave power &
The \(M=64\)-D beam-power vector is converted to
\(\mathrm{float32}\), with NaN and Inf entries replaced by zero, and
normalized as
\(10^{(p_{t,m}-\max_j p_{t,j})/10}\).
For \(L<M\), the normalized vector is masked according to
\(\mathbf{o}_t\). \\
\hline
Beam mask &
The binary observation mask
\(\mathbf{o}_t\in\{0,1\}^{64}\) identifies measured beam-power
entries under partial in-band observation. Both the masked power
vector and the mask are supplied to models that use mmWave power
when \(0<L<M\). \\
\hline
Beam labels &
The best-beam label is
\(m_t^\star=\arg\max_{m\in\{1,\ldots,M\}}p_{t,m}\), computed from
the complete measured beam-power vector before observation masking. \\
\hline

\multicolumn{2}{|l|}{\textbf{Predictor Architectures and Parameterization}} \\
\hline
\textbf{Predictor} & \textbf{Architecture / Parameterization} \\
\hline
Persistence &
Non-learned temporal reference. A circular Gaussian posterior with
spread \(\sigma=1.5\) is centered on the most recent observed beam
and propagated over the \(T\)-step forecasting horizon. \\
\hline
Pwr MLP &
The \(H\)-step power history is flattened, with
\(d_{\mathrm{p}}=HM\) for the unmasked input and
\(d_{\mathrm{p}}=2HM\) when the observation mask is concatenated.
The architecture is
\(\mathrm{FC}(d_{\mathrm{p}},256)\rightarrow\mathrm{ReLU}
\rightarrow\mathrm{Dropout}(0.1)
\rightarrow\mathrm{FC}(256,256)\rightarrow\mathrm{ReLU}\).
Separate
\(\mathrm{FC}(256,TM)\) heads produce future beam logits and
power predictions, where \(TM=5\times64=320\). \\
\hline
G+Pwr GRU &
A single-layer GRU processes the \(H\)-step GPS--power sequence.
Its per-step input dimension is
\(d_{\mathrm{gp}}=4+M\) for full power observation and
\(d_{\mathrm{gp}}=4+2M\) when the mask is concatenated. The GRU has
hidden size \(128\); its final state is processed by
\(\mathrm{FC}(128,128)\rightarrow\mathrm{ReLU}\), followed by
separate \(\mathrm{FC}(128,320)\) beam-logit and power-prediction
heads. \\
\hline
G+Pwr LSTM &
A single-layer LSTM processes the \(H\)-step GPS--power sequence
with per-step input dimension \(d_{\mathrm{gp}}\) and hidden size
\(128\). The final recurrent state is processed by
\(\mathrm{FC}(128,128)\rightarrow\mathrm{ReLU}\), followed by
separate \(\mathrm{FC}(128,320)\) beam-logit and power-prediction
heads. \\
\hline
G+Pwr CNN &
The GPS--power sequence is projected to \(256\) temporal channels
by a one-dimensional input projection and then processed by three
\(\mathrm{Conv1d}(256,256,3,1)\) temporal blocks with ReLU
activation and dropout \(0.1\). The resulting temporal
representation is mapped through separate \(320\)-D beam-logit and
power-prediction heads. \\
\hline
Sensor CNN &
Camera, radar, and LiDAR are processed by separate spatial branches:
\(\mathrm{Conv2d}(C_{\mathrm{in}},8,3,2,1)
\rightarrow\mathrm{ReLU}
\rightarrow\mathrm{Conv2d}(8,16,3,2,1)
\rightarrow\mathrm{ReLU}
\rightarrow\mathrm{AdaptiveAvgPool2d}(1,1)\),
where \(C_{\mathrm{in}}\in\{3,4,1\}\), respectively. GPS is encoded
by
\(\mathrm{FC}(4,16)\rightarrow\mathrm{ReLU}
\rightarrow\mathrm{FC}(16,16)\rightarrow\mathrm{ReLU}\).
The four \(16\)-D features are concatenated into a \(64\)-D temporal
feature and processed by
\(\mathrm{Conv1d}(64,128,3,1)\) and
\(\mathrm{Conv1d}(128,128,3,1)\), followed by
\(\mathrm{FC}(128,128)\) and separate
\(\mathrm{FC}(128,320)\) beam-logit and power-prediction heads. \\
\hline
Vision-position proxy &
The C+G sensing-aided proxy uses the BeamGuard forecaster with only
camera and GPS encoders active and \(L=0\). It uses the same fusion,
history, autoregressive-decoder, split, horizon, calibration, and
output-head settings as BeamGuard. It represents a same-protocol
vision-position setting rather than an exact reimplementation of a
specific external method. \\
\hline
LiDAR-position proxy &
The Lid+G sensing-aided proxy uses the BeamGuard forecaster with only
LiDAR and GPS encoders active and \(L=0\). It uses the same fusion,
history, autoregressive-decoder, split, horizon, calibration, and
output-head settings as BeamGuard. It represents a same-protocol
LiDAR-position setting rather than an exact reimplementation of a
specific external method. \\
\hline
BeamGuard &
Camera, radar, and LiDAR are encoded using the reported pretrained
ResNet-18 backbones, with input channels \(3\), \(4\), and \(1\),
respectively. Each pooled \(512\)-D backbone feature is projected by
\(\mathrm{FC}(512,256)\). GPS and the power--mask vector are encoded
by two-layer vector encoders that project their inputs through a
\(128\)-D hidden layer to the common dimension \(d=256\), with
dropout \(0.1\). Per-time-step modality fusion uses \(2\) transformer
encoder layers; temporal-history encoding uses \(2\) transformer
encoder layers; and autoregressive forecasting uses \(2\) transformer
decoder layers. Each transformer block has \(4\) attention heads,
feed-forward dimension \(4d=1024\), GELU activation, dropout \(0.1\),
and layer normalization. At each future step, separate
\(\mathrm{FC}(256,64)\) heads produce the beam logits and auxiliary
power prediction. \\
\hline
\end{tabular}

\vspace{2pt}
\begin{minipage}{0.97\textwidth}
\scriptsize
\textit{Note:}
\(\mathrm{FC}(a,b)\) denotes a fully connected layer with input
dimension \(a\) and output dimension \(b\).
\(\mathrm{Conv1d}(a,b,k,p)\) denotes a one-dimensional convolution
with \(a\) input channels, \(b\) output channels, kernel size \(k\),
and padding \(p\).
\(\mathrm{Conv2d}(a,b,k,s,p)\) additionally specifies stride \(s\).
All learned predictors use the same segment-level data partitions,
history length \(H=8\), prediction horizon \(T=5\), and
\(M=64\) beam target. The flattened baseline heads contain
\(TM=320\) outputs and are reshaped to \(T\times M\); BeamGuard
instead produces an \(M\)-dimensional output at each autoregressive
future step. C, R, Lid, G, and Pwr denote camera, radar, LiDAR, GPS,
and mmWave power, respectively.
\end{minipage}
\end{table*}

Table~\ref{tab:implementation_settings} summarizes the main implementation settings. Samples are spaced at a 100 ms cadence. The model uses a history of \(H=8\) samples, corresponding to 0.8 s of past observations, and predicts \(T=5\) future samples, corresponding to a 0.5 s forecasting horizon. The beam-observation budgets, candidate-center setting, and virtual beamwidth set are kept fixed unless explicitly varied in the ablation studies.

\begin{table}[t]
\caption{Main Implementation Settings Used in the Reported Experiments}
\label{tab:implementation_settings}
\centering
\setlength{\tabcolsep}{3pt}
\renewcommand{\arraystretch}{1.06}
\begin{tabular}{|p{100pt}|p{110pt}|}
\hline
\textbf{Parameter} & \textbf{Value} \\
\hline
Codebook size, \(M\) & 64 beams \\
\hline
Sampling cadence & 100 ms \\
\hline
History length, \(H\) & 8 samples \\
\hline
Prediction horizon, \(T\) & 5 samples \\
\hline
Beam-observation budgets, \(L\) & \(\{0,8,16,32,64\}\) \\
\hline
Virtual beamwidth set, \(\mathcal{W}\) & \(\{1,3,5\}\) \\
\hline
Planner candidate centers, \(K_c\) & 10 \\
\hline
Optimizer & AdamW \\
\hline
Initial Learning rate & \(1\times10^{-4}\) \\
\hline
Learning-rate schedule &
Cosine annealing with \(T_{\max}=12\), updated once per epoch;
no warm-up \\
\hline
Weight decay & \(1\times10^{-4}\) \\
\hline
Maximum epochs & 12 \\
\hline
Early-stopping patience & 4 epochs \\
\hline
Training batch size & 2 \\
\hline
Evaluation batch size & 1 \\
\hline
Gradient accumulation & None \\
\hline
Random seeds & \(7,13,23,37,53\) for main sweeps; \(7,13,23\) for transfer/few-shot \\
\hline
\end{tabular}
\end{table}
All learned predictors use a training batch size of \(2\), fixed before the experimental campaign and retained across all seeds, modality configurations, and ablations to maintain a common optimization protocol. A one-epoch full-hybrid audit confirmed that batch size \(4\) is also executable on the \(16\)-GB Apple M2 Pro platform and reduces the audited epoch time, but this feasibility test does not establish equivalent final convergence or calibrated test performance. Batch size \(2\) is therefore reported as the fixed setting used for all results rather than as a hardware-imposed maximum.

\begin{table}[t]
\caption{Model, Loss, Calibration, and Planner Settings}
\label{tab:model_planner_settings}
\centering
\setlength{\tabcolsep}{3pt}
\renewcommand{\arraystretch}{1.06}
\begin{tabular}{|p{92pt}|p{118pt}|}
\hline
\textbf{Component} & \textbf{Setting} \\
\hline
Embedding size, \(d\) & 256 \\
\hline
Fusion encoder & 2 transformer layers, 4 heads \\
\hline
History encoder & 2 transformer layers, 4 heads \\
\hline
Autoregressive decoder & 2 transformer layers, 4 heads \\
\hline
Feed-forward size & \(4d=1024\) \\
\hline
Dropout & 0.1 \\
\hline
Decoder outputs & Beam logits and 64-D power vector \\
\hline
Beam loss & Cross-entropy, \(\mathcal{L}_{\mathrm{ce}}\) \\
\hline
Power-loss weight, \(\alpha_p\) & 1.0 \\
\hline
Distance-loss weight, \(\alpha_d\) & 0.25 \\
\hline
Distance-loss scale, \(\sigma\) & 2.0 \\
\hline
Calibration & Temperature scaling on calibration split \\
\hline
Posterior selection & Validation-based raw/calibrated selection \\
\hline
Candidate-center source & Top posterior and predicted-power rankings \\
\hline
Gain factor, \(\eta(w)\) & \(\eta(1)=1.00,\eta(3)=0.95,\eta(5)=0.90\) \\
\hline
Width penalty & \(w-1\) \\
\hline
Switching weight, \(\lambda_{\mathrm{sw}}\) & 0.02 \\
\hline
Width weight, \(\lambda_w\) & 0.01 \\
\hline
Risk weight, \(\lambda_{\rho}\) & 0.50 \\
\hline
Default risk budget, \(\beta\) & 0.10 \\
\hline
Risk sweep, \(\beta\) & \(\{0.05,0.10,0.15\}\) \\
\hline
Risk-screen threshold, \(\rho_{\max}\) & 0.50 \\
\hline
Outage threshold, \(\gamma\) & 0.40 of oracle power \\
\hline
\end{tabular}
\end{table}

The configuration in Table~\ref{tab:model_planner_settings} is fixed during test evaluation. Only the operational risk budget \(\beta\) is varied in the risk-budget sweep. The planner candidate-center count is fixed at \(K_c=10\) as a complexity--coverage tradeoff: it retains the dominant posterior-ranked and predicted-power-ranked beam hypotheses while avoiding exhaustive all-center search over the 64-beam codebook. The candidate set is formed from the unique union of top posterior-ranked and top predicted-power-ranked centers.
\vspace{-2mm}
\subsection{Evaluation protocols}

The evaluation is organized around four protocol groups. First, the \emph{joint day--night} protocol trains on both Scenarios 32 and 33 and evaluates on held-out segments from the combined set. This protocol measures the main operating performance when both daytime and nighttime data are available during training. Second, the \emph{in-domain} protocols train and test within the same scenario, using either Scenario 32 or Scenario 33. Third, the \emph{zero-shot transfer} protocols train on one scenario and evaluate on the other, thereby measuring day--night generalization without target-domain adaptation. Fourth, the \emph{few-shot transfer} protocols adapt the model using a limited fraction of target-domain data and evaluate whether small amounts of target supervision improve transfer. The transfer/few-shot block is evaluated over three seeds, while the main joint, regime, beam-budget, and risk-budget sweeps use five seeds (see Table~\ref{tab:implementation_settings}).
To assess generalization beyond the original S32/S33 day--night pair, additional held-out scenario tests are conducted on DeepSense-6G Scenarios 31 and 34. The source training, validation, and calibration partitions are drawn from Scenarios 32 and 33, while Scenario 31 or Scenario 34 is used only for testing. Scenario 31 provides an additional daytime road setting, and Scenario 34 provides a different nighttime crossroad setting \cite{b26, b27}. These experiments keep the preprocessing, history length, prediction horizon, beam codebook, planner settings, and evaluation metrics unchanged.

These protocols reflect practical deployment requirements by evaluating performance with representative training data, robustness under domain shift, and adaptation using limited target-domain data without training a new model from scratch.
\vspace{-2mm}
\subsection{Operating-Regime and Ablation design}

The experiments instantiate the operating modes defined in Sec.~\ref{sec:beamguard_method}, to isolate the roles of exteroceptive sensing, communication-side measurements, and multimodal fusion. The evaluated regimes include sensor-only operation using camera--radar--LiDAR--GPS, power-only operation using mmWave beam-power observations, lightweight hybrid GPS--power operation, geometry-assisted GPS--LiDAR--power operation, and full camera--radar--LiDAR--GPS--power fusion. This focused ablation design captures deployment-relevant operating points without exhaustively enumerating every modality subset.

 The beam-observation budget ablation varies the number of measured entries in the mmWave power vector using the values in Table~\ref{tab:implementation_settings}. For partial in-band observation, the \(L\) observed beam entries are selected using deterministic uniform subsampling over the \(M=64\) beam codebook. The corresponding binary mask \(\mathbf{o}_t\) is provided to the forecaster together with the masked power vector, allowing unobserved entries to be distinguished from measured low-power values. The mask is fixed for each \(L\), so the budget sweep isolates the effect of measurement availability from that of adaptive probing. The cases \(L=8\), \(16\), and \(32\) represent progressively reduced beam-observation budgets, \(L=64\) denotes full power-vector observation, and \(L=0\) provides the sensor-only reference.

Sensitivity to the mask-generation is evaluated using an alternative non-learned local-neighborhood probing policy for \(L\in\{8,16,32\}\). The forecasting architecture, data partitions, beam-observation budgets, planner settings, and random seeds are held fixed, and only the mask-generation rule used to produce \(\mathbf{o}_t\) is changed. This comparison determines whether the partial-observation results depend strongly on deterministic uniform subsampling. Learned probing is not included because it would introduce a separate sequential policy-optimization problem beyond the present mask-aware forecasting-and-control framework.

Risk-budget ablations vary the operational budget \(\beta\) in \eqref{eq:planner_score} to evaluate how planner conservativeness affects the resulting outage--gain--switching tradeoff. This analysis is conducted for the G+Lid+Pwr and full-hybrid configurations, which represent the principal geometry-assisted and complete multimodal operating regimes.

Two complementary planner ablations isolate the effects of virtual beamwidth and controller decision rule. First, fixed-width runs set the allowable width set \(\mathcal{W}_{\mathrm{eval}}\) to \(\{1\}\), \(\{3\}\), or \(\{5\}\), while the adaptive planner uses \(\{1,3,5\}\). These variants use the same trained full-hybrid forecaster, selected beam posterior, auxiliary power prediction, candidate-center construction, test windows, and planner parameters; only the allowable virtual beamwidth set is changed. This comparison separates the reliability effect of covering more adjacent codebook beams from adaptive width selection.
Second, a same-forecaster controller ablation isolates the contribution of the controller decision rule. The trained checkpoint, selected beam posterior, auxiliary power prediction, and \(940\) test windows are identical across all variants. The comparison includes greedy maximum-posterior selection with fixed \(w=1\) and \(w=3\), a widest-safe heuristic, a risk-ablated adaptive controller, and the complete risk-aware BeamGuard planner. No additional controller is learned or retrained. Consequently, differences in threshold-based outage, covered-power gain ratio, and switching rate arise from controller logic rather than predictor performance.
\vspace{-2mm}
\subsection{Baselines}

The evaluation uses controlled predictor and controller baselines to isolate the effects of temporal modeling, sensing information, in-band observations, multimodal fusion, and planner logic. All learned predictor baselines use the same segment-level train, validation, calibration, and test partitions, history length \(H=8\), prediction horizon \(T=5\), and \(M=64\) beam-label target as BeamGuard. The non-learned persistence baseline is evaluated on the same test windows. For models that use partial in-band observations, the masked power vector and its binary observation mask are provided consistently so that unobserved beam entries are not interpreted as measured low-power values.

The predictor set includes last-beam persistence, a power-only multilayer perceptron (MLP), G+Pwr recurrent models based on gated recurrent units (GRU) and long short-term memory (LSTM) units, a G+Pwr temporal convolutional neural network (CNN), a sensor-only temporal CNN, and the BeamGuard transformer-style forecaster. The persistence and MLP baselines evaluate short-term memory and static in-band prediction, respectively, while the GRU, LSTM, and temporal CNN models evaluate recurrent and convolutional temporal aggregation. BeamGuard evaluates transformer-based multimodal fusion, temporal encoding, autoregressive forecasting, and auxiliary power prediction under the same data protocol.

Same-protocol C+G and Lid+G baselines are additionally included to represent vision-position and LiDAR-position sensing-aided beam-prediction settings. These proxy baselines are evaluated using the same splits, forecasting horizon, codebook target, calibration procedure, and metrics as BeamGuard. They are not presented as exact re-implementations of any individual published method; rather, they provide controlled numerical references for the sensing-aided modality families discussed in the related work. Table~\ref{tab:preprocess_baseline_summary} summarizes the preprocessing and baseline architectures, while their numerical results are reported in Table~\ref{tab:baselines}.

Recent transformer-based and multimodal beam-prediction methods~\cite{b23a,b23b,b23c} provide useful external context for BeamGuard. Their published numerical results are not inserted as direct baselines because the corresponding protocols generally differ in prediction horizon, modality availability, train/test partitioning, beam-observation assumptions, calibration treatment, and the inclusion of downstream beam-control metrics. BeamGuard evaluates multi-step future beam posteriors, partial in-band beam-observation budgets, calibration-aware posterior selection, and risk-aware virtual beamwidth control using threshold-based outage, gain ratio, and switching behavior. Numerical comparisons are therefore restricted to models evaluated under the common BeamGuard protocol.

Planner-side performance is evaluated through the risk-budget sweep, fixed-width ablation, and a same-forecaster controller comparison. The fixed-width experiment evaluates \(w\in\{1,3,5\}\) to isolate the effect of adjacent-beam coverage. The controller comparison includes greedy maximum-posterior selection with fixed \(w=1\) and \(w=3\), a widest-safe heuristic, a risk-ablated adaptive controller, and the complete risk-aware BeamGuard planner. All controller variants use the same trained checkpoint, selected beam posterior, auxiliary power prediction, and test windows; only the controller decision rule is changed. This design isolates controller-side behavior from predictor performance without introducing an additional learned control policy.
\vspace{-2mm}
\subsection{Evaluation Metrics}

The predictor is evaluated using ranked beam accuracy, probabilistic calibration, beam-index agreement, and auxiliary power-regression metrics. For \(K_{\mathrm{acc}}\in\{1,3,5\}\), top-\(K_{\mathrm{acc}}\) accuracy measures whether the ground-truth best beam appears among the \(K_{\mathrm{acc}}\) highest-probability entries of the predicted posterior. Here, \(K_{\mathrm{acc}}\) is used only as an evaluation index for ranked prediction accuracy and is distinct from \(K_c\), which controls the number of candidate beam centers considered by the planner. The metric is defined as
\begin{equation}
    \mathrm{Acc}@K_{\mathrm{acc}}
    =
    \frac{1}{N_{\mathrm{f}}}
    \sum_{n=1}^{N_{\mathrm{f}}}
    \mathbf{1}
    \left\{
    m_n^\star
    \in
    \mathrm{Top}_{K_{\mathrm{acc}}}(\widehat{\mathbf{q}}_n)
    \right\},
    \label{eq:topk_metric}
\end{equation}
where \(N_{\mathrm{f}}\) is the number of evaluated future-step samples, \(m_n^\star\) is the corresponding best-beam label, \(\widehat{\mathbf{q}}_n\) is the predicted beam posterior, and \(\mathrm{Top}_{K_{\mathrm{acc}}}(\widehat{\mathbf{q}}_n)\) returns the indices of the \(K_{\mathrm{acc}}\) largest posterior entries.

NLL evaluates the probability assigned to the correct beam and penalizes overconfident wrong predictions. The Brier score measures the squared error between the predicted beam posterior and the one-hot beam label. ECE measures the mismatch between predicted confidence and empirical correctness, which is important because the planner uses posterior mass to estimate beam coverage and miscoverage risk. Beam-distance agreement, reported as DBA@3, measures whether the predicted beam lies within three beam indices of the ground-truth best beam on the ordered codebook. Auxiliary power mean absolute error, denoted Power MAE, measures the accuracy of the predicted beam-power vector. Together, these metrics evaluate not only whether the most likely beam is correct, but also whether the predicted distribution and auxiliary power structure are useful for risk-aware beam planning (see Table~\ref{tab:budget_diagnostics}).

The planner is evaluated using communication-level metrics. Let
\begin{equation}
    p_n^{\mathrm{sel}}
    =
    \max_{m\in\mathcal{V}(c_n,w_n)} p_{n,m},
    \qquad
    p_n^{\mathrm{orc}}
    =
    \max_{m\in\{1,\ldots,M\}} p_{n,m},
    \label{eq:selected_oracle_power}
\end{equation}
denote the best received power covered by the selected virtual beam and the oracle best narrow-beam power, respectively, for evaluated action \(n\). For a threshold \(\gamma\in(0,1)\), the threshold-based empirical outage probability is
\begin{equation}
    P_{\mathrm{out}}(\gamma)
    =
    \frac{1}{N_{\mathrm{a}}}
    \sum_{n=1}^{N_{\mathrm{a}}}
    \mathbf{1}
    \left\{
    p_n^{\mathrm{sel}} < \gamma p_n^{\mathrm{orc}}
    \right\},
    \label{eq:empirical_outage}
\end{equation}
where \(N_{\mathrm{a}}\) is the number of evaluated planner actions. In the experiments, \(\gamma=0.40\). Thus, \(P_{\mathrm{out}}(\gamma)\) measures the fraction of decisions for which the selected virtual beam fails to retain at least \(40\%\) of the oracle best-beam received power. This metric is distinct from the predicted beam-miscoverage risk \(\rho_t(c,w)\) used inside the planner score.

The gain ratio is computed as
\begin{equation}
    R_{\mathrm{gain}}
    =
    \frac{1}{N_{\mathrm{a}}}
    \sum_{n=1}^{N_{\mathrm{a}}}
    \frac{p_n^{\mathrm{sel}}}{p_n^{\mathrm{orc}}},
    \label{eq:gain_ratio}
\end{equation}
which measures the fraction of oracle narrow-beam power retained by the selected virtual beam. A value close to one indicates that the selected action preserves most of the best achievable beamforming gain. 

The switching rate is
\begin{equation}
    R_{\mathrm{sw}}
    =
    \frac{1}{N_{\mathrm{a}}-1}
    \sum_{n=2}^{N_{\mathrm{a}}}
    \mathbf{1}\{c_n\neq c_{n-1}\}.
    \label{eq:switch_rate}
\end{equation}
And it measures how frequently the selected beam center changes over time. Together, \(P_{\mathrm{out}}(\gamma)\), \(R_{\mathrm{gain}}\), and \(R_{\mathrm{sw}}\) quantify the reliability, efficiency, and temporal stability of the beam-control policy.

All reported multi-seed results are aggregated using the mean and standard deviation across random seeds. 
All experiments use the same preprocessing, split construction, model configuration, and planner settings unless otherwise stated. Hyperparameters are selected using the validation split, calibration parameters are fitted only on the calibration split, and final performance is reported on the held-out test split. This separation prevents test-set leakage and ensures that the reported predictor and planner metrics reflect out-of-sample performance.
\vspace{-2mm}
\subsection{Reproducibility and Code Availability}

The BeamGuard implementation, experiment configurations, preprocessing and standardization scripts, manifest-generation and audit utilities, segment-level split procedures, training and calibration code, planner and controller evaluation scripts, and figure- and table-generation scripts are publicly available at \url{https://github.com/AbidemiMatthew/BeamGuard}. The exact repository revision aligned with this manuscript and the reported experiments is identified by commit \texttt{373470cd91ed63332ab983648e55c15a2a20a566}.
The repository also provides the Python environment specification, reproduction documentation, citation metadata, and safe manifest templates. The raw DeepSense 6G measurements and trained checkpoints are not redistributed; the raw measurements remain available through the official dataset source \cite{b23}, and the released scripts reconstruct the standardized modality representations, observation masks, manifests, and experimental partitions used in this work.
\vspace{-4mm}
\section{Results and Discussion}
\label{sec:resudisc}
The results evaluate BeamGuard as an end-to-end beam-management framework that connects future beam forecasting with risk-aware adaptive beamwidth control. The discussion therefore emphasizes both prediction quality and communication-level behavior, including threshold-based outage probability, gain ratio, switching behavior, beam-observation budget, transfer robustness, and practical runtime.

\begin{table}[t]
\caption{Anchor Full-Hybrid Performance Under the Joint Day--Night Protocol}
\label{tab:anchor}
\centering
\setlength{\tabcolsep}{3pt}

\begin{tabular}{|p{115pt}|p{95pt}|}
\hline
\textbf{Metric} & \textbf{Value} \\
\hline
Top-1 / Top-3 / Top-5 & 0.393 / 0.778 / 0.897 \\
\hline
NLL / ECE (selected posterior) & 1.837 / 0.031 \\
\hline
Outage probability, \(P_{\mathrm{out}}(0.40)\) & 0.0060 \\
\hline
Gain ratio, \(R_{\mathrm{gain}}\) & 0.895 \\
\hline
Switching rate, \(R_{\mathrm{sw}}\) & 0.134 \\
\hline
Predictor / planner latency & 76.37 / 1.49 ms \\
\hline
\end{tabular}

\vspace{2pt}
\par\noindent
\begin{minipage}{\linewidth}
\raggedright
\footnotesize
\textit{Note:}This table reports the single full-hybrid anchor checkpoint used as the
complete-system reference in the abstract. Multi-seed budget-sweep means are
reported separately in the beam-budget analysis.
\end{minipage}
\end{table}
\subsection{Anchor Hybrid Performance}

The full-hybrid configuration provides the main reference point for the proposed framework because it evaluates the complete BeamGuard pipeline. It uses camera, radar, LiDAR, GPS, and the full \(M=64\) mmWave power vector under the joint day--night protocol. The anchor result is used as the complete-system reference because it combines full multimodal forecasting with planner-level reliability evaluation. In this pipeline, the temporal multimodal forecaster estimates future beam distributions, and the risk-aware planner converts the selected posterior into a beam center and virtual beamwidth according to \eqref{eq:planner_decision}.

Table~\ref{tab:anchor} summarizes the anchor result. The predictor achieves Top-1, Top-3, and Top-5 accuracies of approximately \(0.393\), \(0.778\), and \(0.897\), respectively. These values are reported as the full-hybrid BeamGuard reference, while the G+Pwr result in Table~\ref{tab:baselines} is discussed separately as a same-input predictor baseline. The high Top-5 value is important because the planner is not restricted to a single narrow beam; it uses posterior mass over neighboring beams to estimate coverage probability in \eqref{eq:coverage_prob}. The NLL reflects the likelihood assigned to the correct beam labels, while ECE quantifies the mismatch between predicted confidence and empirical correctness. The threshold-based outage probability \(P_{\mathrm{out}}(0.40)=0.0060\) and gain ratio \(R_{\mathrm{gain}}=0.895\) indicate that the predicted posterior is useful for communication control, not only for beam classification. The latency values further show that the planner adds little computational overhead relative to the forecaster.

\begin{figure*}[t]
\centering
\includegraphics[width=1.7\columnwidth]{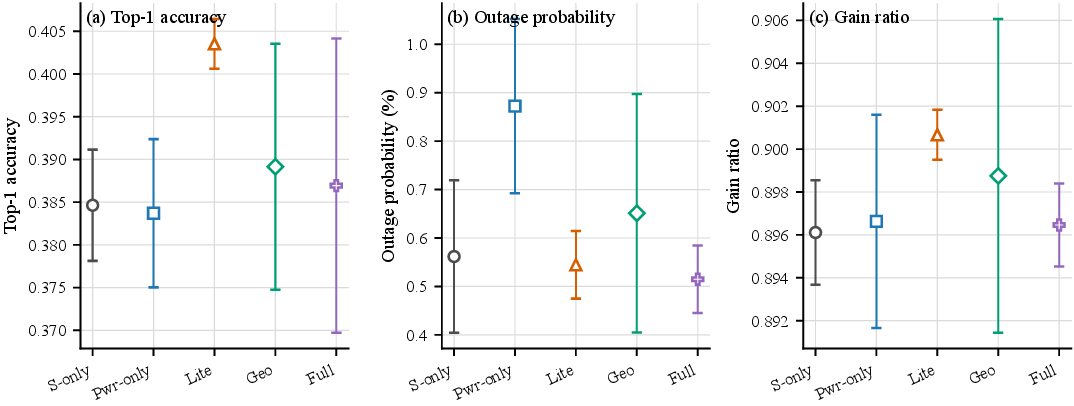}
\vspace{-2mm}
\caption{Operating-regime comparison under joint day--night training. S-only denotes camera--radar--LiDAR--GPS sensing without in-band beam observation; Pwr-only denotes mmWave power only; Lite denotes GPS--mmWave power; Geo denotes GPS--LiDAR--mmWave power; and Full denotes camera--radar--LiDAR--GPS--mmWave power. Here, C, R, Lid, G, and Pwr denote camera, radar, LiDAR, GPS, and mmWave power, respectively. Error bars show standard deviation across five seeds.}
\label{fig:regime_metrics}
\end{figure*}

\subsection{Operating Regimes}

The operating-regime comparison in Fig.~\ref{fig:regime_metrics} summarizes forecasting accuracy, threshold-based outage probability, and gain ratio across sensor-only, in-band-only, and hybrid configurations. The sensor-only configuration uses camera, radar, LiDAR, and GPS without in-band beam-power observations and therefore represents the lowest beam-training-overhead mode. The Pwr-only configuration isolates the value of direct communication-side measurements, while the Lite, Geo, and Full configurations progressively combine mobility, geometric, and multimodal sensing information with in-band power observations.

The results in Fig.~\ref{fig:regime_metrics} show that beam-management quality cannot be judged from top-\(K_{\mathrm{acc}}\) accuracy alone. The Lite configuration achieves strong ranked prediction because GPS and mmWave power provide compact mobility and direct communication-side evidence. In contrast, the Full configuration represents the most reliability-oriented operating point in the regime comparison, achieving the lowest threshold-based empirical outage even though it does not uniformly maximize Top-\(K_{\mathrm{acc}}\) accuracy. This distinction reinforces the BeamGuard design objective of assessing beam forecasting together with its communication-level consequences, including \(P_{\mathrm{out}}(\gamma)\), \(R_{\mathrm{gain}}\), and beam-center switching behavior.

\begin{figure*}[t]
\centering
\includegraphics[width=1.7\columnwidth]{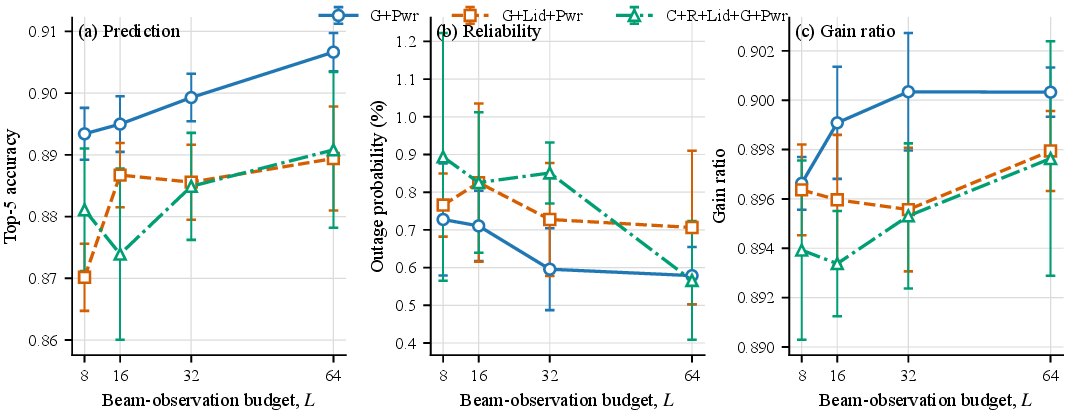}
\vspace{-2mm}
\caption{Beam-observation budget sweep under joint day--night training. The codebook size remains fixed at \(M=64\), while \(L\) denotes the number of observed entries in the mmWave power vector. G+Pwr, G+Lid+Pwr, and C+R+Lid+G+Pwr denote GPS--mmWave power, GPS--LiDAR--mmWave power, and full multimodal hybrid sensing, respectively. Error bars show standard deviation across five seeds.}
\label{fig:budget_sweep}
\end{figure*}

\subsection{Beam-Observation Budget}

The impact of the beam-observation budget \(L\) on ranked prediction, outage probability, and gain ratio is shown in Fig.~\ref{fig:budget_sweep}. The codebook size remains fixed at \(M=64\), while \(L\) controls how many entries of the mmWave power vector are observed. This directly evaluates the partial in-band sensing formulation in \eqref{eq:masked_power}. Increasing \(L\) generally improves ranked prediction and reduces outage, but partial budgets already retain much of the full-budget behavior.

This observation result is important for practical V2I deployment because full beam sweeps provide rich communication-side information but increase training overhead, while sensor-only operation avoids in-band measurement cost but may be more sensitive to sensing-domain variation. BeamGuard supports the intermediate regime by using the observation mask \(\mathbf{o}_t\) to distinguish unmeasured beams from measured low-power beams. 

To complement the Top-\(K_{\mathrm{acc}}\), outage, and gain trends in Fig.~\ref{fig:budget_sweep}, Table~\ref{tab:budget_diagnostics} reports additional distribution-level predictor diagnostics across the beam-observation budgets. The \(L=0\) row corresponds to the sensor-only reference, while \(L=8,16,32,64\) correspond to partial or full in-band power observation. The diagnostics show that larger observed in-band budgets improve the probabilistic quality of the GPS+Pwr predictor. For example, the Brier score decreases from \(0.7465\) at \(L=8\) to \(0.7320\) at \(L=64\), while DBA@3 increases from \(0.9046\) to \(0.9149\). Power MAE also decreases slightly with larger \(L\), indicating that additional in-band observations improve the auxiliary power-forecasting task.

The dependence on the mask-generation rule is evaluated in Table~\ref{tab:sampling_policy_sensitivity}. This sensitivity test compares deterministic uniform subsampling with a local neighborhood probing policy under the same beam-observation budgets, data splits, model configuration, random seeds, and planner settings. 
The results show that BeamGuard's partial-observation behavior is not tied to a single uniform mask pattern. Across the tested regimes and budgets, uniform and local probing produce comparable Top-1/Top-3/Top-5 accuracy, threshold-based outage, gain ratio, and switching behavior. Local probing is slightly better in some cases, while uniform probing is marginally better in others, and no policy uniformly dominates across all regimes and budgets. This supports the interpretation that the main contribution is mask-aware forecasting and risk-aware control under partial in-band observation, while optimized learned probing remains a separate policy-design problem for future work.

\begin{table*}[t]
\centering
\caption{Additional Predictor Diagnostics Across Beam-Observation Budgets}
\label{tab:budget_diagnostics}
\footnotesize
\setlength{\tabcolsep}{5.0pt}
\begin{tabular}{|l|c|c|c|c|}
\hline
\textbf{Regime} & \(\boldsymbol{L}\) & \textbf{Brier \(\downarrow\)} & \textbf{DBA@3 \(\uparrow\)} & \textbf{Power MAE \(\downarrow\)} \\
\hline
C+R+Lid+G & 0 & \(0.7608 \pm 0.0035\) & \(0.8985 \pm 0.0034\) & \(0.1357 \pm 0.0051\) \\
\hline
G+Pwr & 8 & \(0.7465 \pm 0.0054\) & \(0.9046 \pm 0.0023\) & \(0.1163 \pm 0.0009\) \\
\hline
G+Lid+Pwr & 8 & \(0.7724 \pm 0.0046\) & \(0.8924 \pm 0.0017\) & \(0.1234 \pm 0.0024\) \\
\hline
C+R+Lid+G+Pwr & 8 & \(0.7637 \pm 0.0088\) & \(0.8929 \pm 0.0091\) & \(0.1232 \pm 0.0032\) \\
\hline
G+Pwr & 16 & \(0.7419 \pm 0.0051\) & \(0.9075 \pm 0.0031\) & \(0.1161 \pm 0.0014\) \\
\hline
G+Lid+Pwr & 16 & \(0.7590 \pm 0.0077\) & \(0.8990 \pm 0.0049\) & \(0.1190 \pm 0.0019\) \\
\hline
C+R+Lid+G+Pwr & 16 & \(0.7705 \pm 0.0104\) & \(0.8911 \pm 0.0082\) & \(0.1255 \pm 0.0051\) \\
\hline
G+Pwr & 32 & \(0.7371 \pm 0.0023\) & \(0.9111 \pm 0.0017\) & \(0.1152 \pm 0.0014\) \\
\hline
G+Lid+Pwr & 32 & \(0.7594 \pm 0.0082\) & \(0.8973 \pm 0.0053\) & \(0.1206 \pm 0.0011\) \\
\hline
C+R+Lid+G+Pwr & 32 & \(0.7567 \pm 0.0079\) & \(0.8980 \pm 0.0047\) & \(0.1211 \pm 0.0017\) \\
\hline
G+Pwr & 64 & \(\mathbf{0.7320 \pm 0.0021}\) & \(\mathbf{0.9149 \pm 0.0025}\) & \(\mathbf{0.1139 \pm 0.0012}\) \\
\hline
G+Lid+Pwr & 64 & \(0.7444 \pm 0.0066\) & \(0.9049 \pm 0.0045\) & \(0.1152 \pm 0.0024\) \\
\hline
C+R+Lid+G+Pwr & 64 & \(0.7433 \pm 0.0116\) & \(0.9065 \pm 0.0090\) & \(0.1161 \pm 0.0043\) \\
\hline
\end{tabular}

\vspace{2pt}
\begin{minipage}{0.97\textwidth}
\footnotesize
\textit{Note:} Values are mean \(\pm\) standard deviation over five random seeds. The \(L=0\) row is the sensor-only reference, while \(L=8,16,32,64\) correspond to partial or full in-band power observation. Lower Brier score and Power MAE are better; higher DBA@3 is better. C, R, Lid, G, and Pwr denote camera, radar, LiDAR, GPS, and mmWave power, respectively.
\end{minipage}
\end{table*}

\begin{figure*}[t]
\centering
\includegraphics[width=1.0\columnwidth]{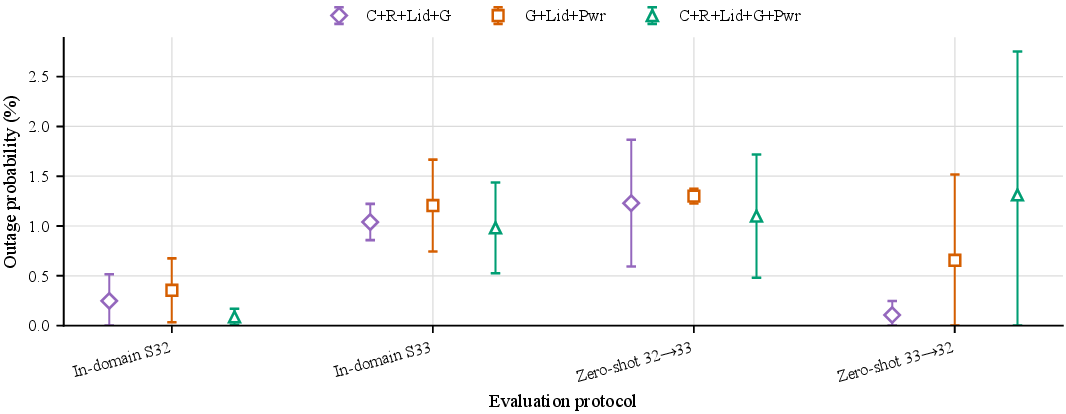}
\caption{Outage probability under in-domain and zero-shot day--night transfer protocols. In-domain S32 and S33 denote training and testing within the same scenario, while zero-shot denotes transfer from the source scenario to the target scenario. C+R+Lid+G, G+Lid+Pwr, and C+R+Lid+G+Pwr denote sensor-only, geometry-assisted hybrid, and full hybrid operation, respectively. Error bars show standard deviation across three seeds.}
\label{fig:protocol_outage}
\end{figure*}

\begin{table*}[!t]
\caption{Sampling-Policy Sensitivity Under Partial In-Band Observation}
\label{tab:sampling_policy_sensitivity}
\centering
\scriptsize
\setlength{\tabcolsep}{2.4pt}
\renewcommand{\arraystretch}{1.08}
\resizebox{\textwidth}{!}{%
\begin{tabular}{|c|c|c|c|c|c|c|c|c|}
\hline
\textbf{Policy} & \textbf{Regime} & \(\boldsymbol{L}\) &
\textbf{Top-1 \(\uparrow\)} &
\textbf{Top-3 \(\uparrow\)} &
\textbf{Top-5 \(\uparrow\)} &
\(\boldsymbol{P_{\mathrm{out}}(0.40)\downarrow}\) &
\(\boldsymbol{R_{\mathrm{gain}}\uparrow}\) &
\(\boldsymbol{R_{\mathrm{sw}}\downarrow}\) \\
\hline
Local & C+R+Lid+G+Pwr & 8 & \(0.3772{\pm}0.0181\) & \(0.7451{\pm}0.0182\) & \(0.8800{\pm}0.0109\) & \(0.0073{\pm}0.0012\) & \(0.8958{\pm}0.0024\) & \(0.1409{\pm}0.0136\) \\
\hline
Uniform & C+R+Lid+G+Pwr & 8 & \(0.3776{\pm}0.0156\) & \(0.7424{\pm}0.0154\) & \(0.8811{\pm}0.0099\) & \(0.0089{\pm}0.0033\) & \(0.8939{\pm}0.0036\) & \(0.1423{\pm}0.0083\) \\
\hline
Local & G+Lid+Pwr & 8 & \(0.3819{\pm}0.0136\) & \(0.7420{\pm}0.0136\) & \(0.8817{\pm}0.0090\) & \(0.0076{\pm}0.0017\) & \(0.8958{\pm}0.0014\) & \(0.1413{\pm}0.0067\) \\
\hline
Uniform & G+Lid+Pwr & 8 & \(0.3751{\pm}0.0050\) & \(0.7328{\pm}0.0025\) & \(0.8702{\pm}0.0054\) & \(0.0077{\pm}0.0008\) & \(0.8964{\pm}0.0018\) & \(0.1417{\pm}0.0074\) \\
\hline
Local & G+Pwr & 8 & \(0.4023{\pm}0.0046\) & \(0.7657{\pm}0.0053\) & \(0.8931{\pm}0.0036\) & \(0.0074{\pm}0.0015\) & \(0.8967{\pm}0.0014\) & \(0.1353{\pm}0.0103\) \\
\hline
Uniform & G+Pwr & 8 & \(0.4010{\pm}0.0052\) & \(0.7665{\pm}0.0058\) & \(0.8934{\pm}0.0042\) & \(0.0073{\pm}0.0015\) & \(0.8966{\pm}0.0011\) & \(0.1382{\pm}0.0093\) \\
\hline
Local & C+R+Lid+G+Pwr & 16 & \(0.3792{\pm}0.0094\) & \(0.7511{\pm}0.0064\) & \(0.8858{\pm}0.0049\) & \(0.0068{\pm}0.0011\) & \(0.8970{\pm}0.0017\) & \(0.1404{\pm}0.0119\) \\
\hline
Uniform & C+R+Lid+G+Pwr & 16 & \(0.3662{\pm}0.0111\) & \(0.7365{\pm}0.0183\) & \(0.8739{\pm}0.0139\) & \(0.0083{\pm}0.0019\) & \(0.8934{\pm}0.0021\) & \(0.1377{\pm}0.0124\) \\
\hline
Local & G+Lid+Pwr & 16 & \(0.3799{\pm}0.0135\) & \(0.7484{\pm}0.0160\) & \(0.8850{\pm}0.0110\) & \(0.0071{\pm}0.0010\) & \(0.8971{\pm}0.0020\) & \(0.1434{\pm}0.0044\) \\
\hline
Uniform & G+Lid+Pwr & 16 & \(0.3834{\pm}0.0060\) & \(0.7539{\pm}0.0122\) & \(0.8867{\pm}0.0052\) & \(0.0083{\pm}0.0021\) & \(0.8960{\pm}0.0026\) & \(0.1411{\pm}0.0050\) \\
\hline
Local & G+Pwr & 16 & \(0.4037{\pm}0.0080\) & \(0.7792{\pm}0.0079\) & \(0.8966{\pm}0.0038\) & \(0.0072{\pm}0.0013\) & \(0.8989{\pm}0.0029\) & \(0.1474{\pm}0.0124\) \\
\hline
Uniform & G+Pwr & 16 & \(0.4043{\pm}0.0062\) & \(0.7781{\pm}0.0092\) & \(0.8950{\pm}0.0045\) & \(0.0071{\pm}0.0009\) & \(0.8991{\pm}0.0023\) & \(0.1504{\pm}0.0069\) \\
\hline
Local & C+R+Lid+G+Pwr & 32 & \(0.3771{\pm}0.0047\) & \(0.7431{\pm}0.0078\) & \(0.8809{\pm}0.0041\) & \(0.0088{\pm}0.0048\) & \(0.8936{\pm}0.0032\) & \(0.1409{\pm}0.0122\) \\
\hline
Uniform & C+R+Lid+G+Pwr & 32 & \(0.3850{\pm}0.0132\) & \(0.7528{\pm}0.0078\) & \(0.8849{\pm}0.0087\) & \(0.0085{\pm}0.0008\) & \(0.8953{\pm}0.0029\) & \(0.1397{\pm}0.0111\) \\
\hline
Local & G+Lid+Pwr & 32 & \(0.3813{\pm}0.0109\) & \(0.7517{\pm}0.0116\) & \(0.8863{\pm}0.0037\) & \(0.0074{\pm}0.0016\) & \(0.8969{\pm}0.0034\) & \(0.1434{\pm}0.0097\) \\
\hline
Uniform & G+Lid+Pwr & 32 & \(0.3831{\pm}0.0109\) & \(0.7505{\pm}0.0145\) & \(0.8856{\pm}0.0061\) & \(0.0073{\pm}0.0015\) & \(0.8956{\pm}0.0025\) & \(0.1420{\pm}0.0067\) \\
\hline
Local & G+Pwr & 32 & \(0.4077{\pm}0.0056\) & \(0.7817{\pm}0.0065\) & \(0.8997{\pm}0.0034\) & \(0.0064{\pm}0.0010\) & \(0.8996{\pm}0.0012\) & \(0.1413{\pm}0.0066\) \\
\hline
Uniform & G+Pwr & 32 & \(0.4102{\pm}0.0035\) & \(0.7826{\pm}0.0042\) & \(0.8993{\pm}0.0038\) & \(0.0060{\pm}0.0011\) & \(0.9003{\pm}0.0024\) & \(0.1421{\pm}0.0064\) \\
\hline
\end{tabular}%
}

\vspace{2pt}
\begin{minipage}{0.97\textwidth}
\footnotesize
\textit{Note:} Values are mean \(\pm\) standard deviation over five random seeds. Uniform denotes deterministic uniformly spaced beam subsampling, whereas Local denotes the alternative neighborhood-style probing rule. All rows use the same data protocol, model class, observation budget, and planner settings; only the mask-generation policy changes. Learned probing is not included because it introduces a separate policy-learning problem.
\end{minipage}
\end{table*}

\subsection{Day-Night Transfer and Few-shot Adaptation}

The outage behavior under in-domain and zero-shot day--night transfer protocols is shown in Fig.~\ref{fig:protocol_outage}. In-domain results are generally more stable because training and testing are drawn from the same scenario distribution. Zero-shot transfer is more challenging because illumination, sensing statistics, and beam-transition patterns differ across the day and night scenarios. The hybrid configurations remain competitive under transfer because they combine complementary sensing and communication evidence. In particular, the geometry-assisted hybrid mode, G+Lid+Pwr, combines coarse mobility, geometric structure, and in-band power measurements, while the full-hybrid mode uses the complete sensing stack.

To substantiate the limited target-domain adaptation claim, Table~\ref{tab:fewshot_transfer} compares zero-shot transfer with \(20\%\) few-shot adaptation for the two main hybrid regimes. The table reports ranked beam-forecasting accuracy together with planner-level outage and gain metrics, allowing the effect of target-domain supervision to be evaluated at both the prediction and communication-control levels.
\begin{table*}[!t]
\caption{Zero-Shot and Few-Shot Transfer Results}
\label{tab:fewshot_transfer}
\centering
\scriptsize
\setlength{\tabcolsep}{3.0pt}
\renewcommand{\arraystretch}{1.12}
\resizebox{\textwidth}{!}{%
\begin{tabular}{|c|c|c|c|c|c|c|c|}
\hline
\textbf{Direction} & \textbf{Regime} & \textbf{Setting} &
\textbf{Top-1 \(\uparrow\)} & \textbf{Top-3 \(\uparrow\)} &
\textbf{Top-5 \(\uparrow\)} &
\(\boldsymbol{P_{\mathrm{out}}(0.40)\downarrow}\) &
\(\boldsymbol{R_{\mathrm{gain}}\uparrow}\) \\
\hline
S32\(\rightarrow\)S33 & G+Lid+Pwr & Zero-shot
& \(0.1499{\pm}0.0316\) & \(0.3602{\pm}0.0352\) & \(0.5544{\pm}0.0188\)
& \(0.0130{\pm}0.0007\) & \(0.8675{\pm}0.0098\) \\
\hline
S32\(\rightarrow\)S33 & G+Lid+Pwr & Few-shot 20\%
& \(0.2569{\pm}0.0204\) & \(0.5358{\pm}0.0362\) & \(0.7045{\pm}0.0546\)
& \(0.0128{\pm}0.0022\) & \(0.8852{\pm}0.0048\) \\
\hline
S32\(\rightarrow\)S33 & C+R+Lid+G+Pwr & Zero-shot
& \(0.1403{\pm}0.0402\) & \(0.3184{\pm}0.0937\) & \(0.5096{\pm}0.1343\)
& \(0.0110{\pm}0.0062\) & \(0.8567{\pm}0.0230\) \\
\hline
S32\(\rightarrow\)S33 & C+R+Lid+G+Pwr & Few-shot 20\%
& \(0.2779{\pm}0.0214\) & \(0.5890{\pm}0.0443\) & \(0.7376{\pm}0.0449\)
& \(0.0184{\pm}0.0055\) & \(0.8820{\pm}0.0060\) \\
\hline
S33\(\rightarrow\)S32 & G+Lid+Pwr & Zero-shot
& \(0.2168{\pm}0.0326\) & \(0.4885{\pm}0.0642\) & \(0.6282{\pm}0.0728\)
& \(0.0066{\pm}0.0086\) & \(0.8418{\pm}0.0111\) \\
\hline
S33\(\rightarrow\)S32 & G+Lid+Pwr & Few-shot 20\%
& \(0.3211{\pm}0.0382\) & \(0.6693{\pm}0.0545\) & \(0.8197{\pm}0.0480\)
& \(0.0020{\pm}0.0025\) & \(0.8884{\pm}0.0130\) \\
\hline
S33\(\rightarrow\)S32 & C+R+Lid+G+Pwr & Zero-shot
& \(0.1933{\pm}0.0062\) & \(0.4252{\pm}0.0131\) & \(0.5436{\pm}0.0235\)
& \(0.0131{\pm}0.0144\) & \(0.8179{\pm}0.0255\) \\
\hline
S33\(\rightarrow\)S32 & C+R+Lid+G+Pwr & Few-shot 20\%
& \(0.3108{\pm}0.0388\) & \(0.6654{\pm}0.0521\) & \(0.8145{\pm}0.0467\)
& \(0.0002{\pm}0.0003\) & \(0.8945{\pm}0.0132\) \\
\hline
\end{tabular}%
}
\vspace{2pt}
\begin{minipage}{0.97\textwidth}
\footnotesize
\textit{Note:} Values are mean \(\pm\) standard deviation over three seeds, \(7,13,23\). Few-shot adaptation uses \(20\%\) target-domain supervision. Top-1/Top-3/Top-5 are computed from the calibrated beam posterior. \(P_{\mathrm{out}}(0.40)\) denotes the threshold-based empirical outage probability using \(40\%\) of oracle best-beam power, and \(R_{\mathrm{gain}}\) denotes the normalized gain ratio. C, R, Lid, G, and Pwr denote camera, radar, LiDAR, GPS, and mmWave power, respectively.
\end{minipage}
\end{table*}

The Table~\ref{tab:fewshot_transfer} show that \(20\%\) target-domain adaptation consistently improves ranked beam forecasting relative to zero-shot transfer. For S32\(\rightarrow\)S33, the full-hybrid configuration improves Top-1/Top-3/Top-5 from \(0.1403/0.3184/0.5096\) to \(0.2779/0.5890/0.7376\). For S33\(\rightarrow\)S32, the same configuration improves Top-1/Top-3/Top-5 from \(0.1933/0.4252/0.5436\) to \(0.3108/0.6654/0.8145\). The G+Lid+Pwr configuration shows the same ranked-prediction trend, indicating that limited target-domain supervision improves transfer behavior even without the full sensing stack. Planner-level outcomes also improve in most transfer settings, especially for S33\(\rightarrow\)S32, although the S32\(\rightarrow\)S33 full-hybrid case shows that better ranked prediction does not always guarantee lower threshold-based outage. These results support few-shot adaptation as a practical mechanism for mitigating day--night domain shift, while also showing that BeamGuard remains sensitive to target-domain variation.

\begin{figure*}[!t]
\centering
\includegraphics[width=0.8\textwidth]{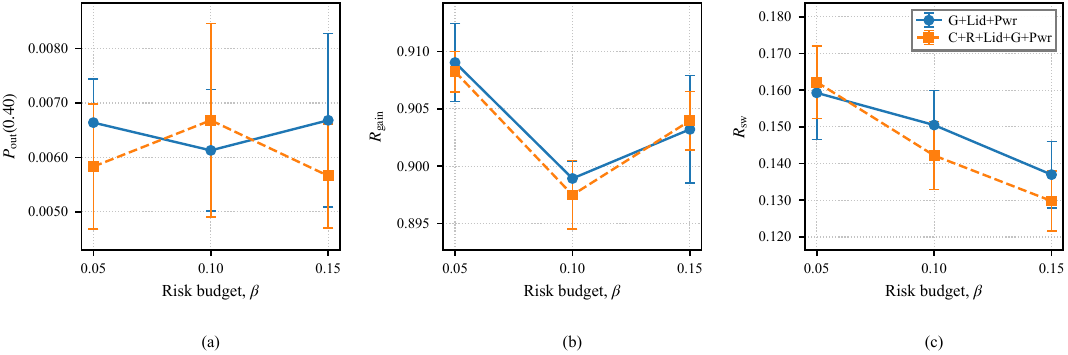}
\caption{Risk-budget sensitivity under the joint day--night protocol for the G+Lid+Pwr and C+R+Lid+G+Pwr configurations. Panels (a), (b), and (c) report the threshold-based empirical outage probability \(P_{\mathrm{out}}(0.40)\), normalized covered-power gain ratio \(R_{\mathrm{gain}}\), and beam-center switching rate \(R_{\mathrm{sw}}\), respectively. The operational risk budget is varied over \(\beta\in\{0.05,0.10,0.15\}\), while the candidate-center construction, allowable virtual beamwidths, hard risk screen, gain term, width penalty, and switching penalty remain fixed. Markers indicate mean values, and error bars represent $\pm$ one standard deviation across five random seeds.}
\label{fig:risk_sweep_outage}
\end{figure*}

\subsection{Additional Held-out Scenario Generalization}

Table~\ref{tab:heldout_scenarios} reports held-out scenario results on Scenarios 31 and 34. This experiment evaluates whether the BeamGuard forecasting-and-control pipeline transfers beyond the original S32/S33 day--night pair. The source model is trained, validated, and calibrated using Scenarios 32 and 33, while Scenario 31 or Scenario 34 is reserved for testing. The experiment is intentionally compact and focuses on the main operating regimes rather than repeating the full ablation suite.

\begin{table*}[!t]
\caption{Held-Out Scenario Stress Test}
\label{tab:heldout_scenarios}
\centering
\scriptsize
\setlength{\tabcolsep}{2.5pt}
\renewcommand{\arraystretch}{1.08}
\resizebox{\textwidth}{!}{%
\begin{tabular}{|c|c|c|c|c|c|c|c|}
\hline
\textbf{Target} & \textbf{Regime} & \(\boldsymbol{L}\) &
\textbf{Top-1 \(\uparrow\)} &
\textbf{Top-3 \(\uparrow\)} &
\textbf{Top-5 \(\uparrow\)} &
\(\boldsymbol{P_{\mathrm{out}}(0.40)\downarrow}\) &
\(\boldsymbol{R_{\mathrm{gain}}\uparrow}\) \\
\hline
S31 & C+R+Lid+G & 0
& \(0.0176{\pm}0.0009\) & \(0.0423{\pm}0.0017\) & \(0.0666{\pm}0.0030\)
& \(0.0542{\pm}0.0021\) & \(0.7270{\pm}0.0017\) \\
\hline
S31 & G+Pwr & 64
& \(0.0453{\pm}0.0069\) & \(0.1181{\pm}0.0148\) & \(0.1718{\pm}0.0221\)
& \(0.0177{\pm}0.0045\) & \(0.7740{\pm}0.0110\) \\
\hline
S31 & G+Lid+Pwr & 64
& \(0.0230{\pm}0.0103\) & \(0.0611{\pm}0.0285\) & \(0.0997{\pm}0.0433\)
& \(0.0347{\pm}0.0111\) & \(0.7447{\pm}0.0143\) \\
\hline
S31 & C+R+Lid+G+Pwr & 64
& \(0.0209{\pm}0.0105\) & \(0.0574{\pm}0.0259\) & \(0.0928{\pm}0.0418\)
& \(0.0409{\pm}0.0117\) & \(0.7401{\pm}0.0148\) \\
\hline
S34 & C+R+Lid+G & 0
& \(0.0071{\pm}0.0047\) & \(0.0213{\pm}0.0116\) & \(0.0337{\pm}0.0167\)
& \(0.3826{\pm}0.0223\) & \(0.4955{\pm}0.0124\) \\
\hline
S34 & G+Pwr & 64
& \(0.0217{\pm}0.0216\) & \(0.0621{\pm}0.0521\) & \(0.0947{\pm}0.0671\)
& \(0.3027{\pm}0.0949\) & \(0.5515{\pm}0.0734\) \\
\hline
S34 & G+Lid+Pwr & 64
& \(0.0383{\pm}0.0186\) & \(0.0950{\pm}0.0453\) & \(0.1439{\pm}0.0675\)
& \(0.3018{\pm}0.0669\) & \(0.5765{\pm}0.0585\) \\
\hline
S34 & C+R+Lid+G+Pwr & 64
& \(0.0593{\pm}0.0418\) & \(0.1477{\pm}0.0884\) & \(0.2166{\pm}0.1232\)
& \(0.2236{\pm}0.0940\) & \(0.6384{\pm}0.0872\) \\
\hline
\end{tabular}%
}
\vspace{2pt}
\begin{minipage}{0.97\textwidth}
\footnotesize
\textit{Note:} Values are mean \(\pm\) standard deviation over five random seeds. Models are trained, validated, and calibrated on Scenarios 32 and 33, while S31 or S34 is used only as held-out testing. \(P_{\mathrm{out}}(0.40)\) is the threshold-based empirical outage probability using \(40\%\) of oracle best-beam power, and \(R_{\mathrm{gain}}\) denotes the normalized covered-power gain ratio.
\end{minipage}
\end{table*}
Table~\ref{tab:heldout_scenarios} shows that held-out scenario transfer is substantially more challenging than the S32/S33 day--night evaluation. The ranked prediction accuracies decrease sharply on both S31 and S34, confirming that changes in road geometry and scenario distribution introduce a stronger domain shift than illumination variation alone. Nevertheless, communication-side observations improve reliability relative to the sensor-only setting. On S31, G+Pwr gives the lowest threshold-based outage and highest gain ratio among the evaluated regimes. On S34, the full-hybrid configuration gives the best ranked prediction, lowest outage, and highest gain ratio among the evaluated regimes. These results strengthen the evaluation by extending BeamGuard beyond the original S32/S33 pair, while also showing that broader scenario generalization remains an open challenge.
\vspace{-2mm}
\subsection{Risk-Budget Sensitivity}

Fig.~\ref{fig:risk_sweep_outage} jointly reports the threshold-based outage probability, covered-power gain ratio, and beam-center switching rate for the G+Lid+Pwr and full-hybrid configurations across \(\beta\in\{0.05,0.10,0.15\}\). The operational budget \(\beta\) controls the excess penalty
\(\lambda_{\rho}[\rho_t(c,w)-\beta]_+\) on the predicted beam-miscoverage risk in \eqref{eq:planner_score}. It does not directly define \(P_{\mathrm{out}}(0.40)\), which is computed from the measured received power covered by the selected action. The configured risk budget and the empirical outage metric are therefore related through the planner decision, but they are not expected to vary proportionally.

For the full-hybrid configuration, \(P_{\mathrm{out}}(0.40)\) remains between \(0.0057\) and \(0.0067\), while \(R_{\mathrm{gain}}\) remains between \(0.8975\) and \(0.9082\). From \(\beta=0.05\) to \(0.15\), the absolute outage change is only \(-1.7\times10^{-4}\), and the gain ratio decreases by \(0.0043\). In contrast, the switching rate decreases from \(0.1622\) to \(0.1298\), corresponding to a \(20.0\%\) reduction. The G+Lid+Pwr configuration exhibits a similar trend. Its endpoint outage changes by only \(4.3\times10^{-5}\), its gain ratio decreases by \(0.0058\), and its switching rate decreases from \(0.1593\) to \(0.1370\), corresponding to a \(14.0\%\) reduction.

The outage and gain values vary only marginally and non-monotonically over the tested range, whereas switching decreases consistently as \(\beta\) increases. The limited variation in outage is consistent with the fixed hard risk screen, the discrete planner action set, and the unchanged gain, beamwidth, and switching terms, while only the soft excess-risk budget \(\beta\) is varied. Relaxing \(\beta\) reduces the excess-risk penalty and can reorder candidate actions with similar posterior coverage and covered-power gain, allowing the switching penalty to favor beam-center continuity without materially changing average outage. The principal visible effect of the tested risk-budget range is therefore a reliability--temporal-stability tradeoff rather than a pronounced outage--gain tradeoff.

These results further demonstrate that the planner operates as a communication-control layer on top of the forecaster. Rather than selecting only the maximum-posterior narrow beam, it uses the predicted beam distribution and auxiliary power estimate to select a beam center and virtual beamwidth while accounting for reliability, gain, width, and switching costs. This supports the treatment of mmWave V2I beam management as a forecasting-and-control problem rather than an accuracy-only prediction task. The aggregate risk-sweep outputs do not record hard-screen or fallback activation frequencies; consequently, the observed trends are not attributed to a measured fallback rate.
\vspace{-2mm}
\subsection{Planner Controller Ablations}

To isolate the effect of adaptive virtual beamwidth selection, Table~\ref{tab:fixed_width_ablation} evaluates fixed-width planner variants on the same predicted beam posterior. In these ablations, the trained forecaster, test split, candidate-center generation, and posterior inputs are unchanged; only the allowable virtual beamwidth set is fixed to \(\{1\}\), \(\{3\}\), or \(\{5\}\). This separates the effect of adaptive width selection from the reliability gain that can arise simply from covering more adjacent beams.

\begin{table}[!t]
\caption{Fixed-Width Planner Ablation Using Identical Forecast Outputs}
\label{tab:fixed_width_ablation}
\centering
\footnotesize
\setlength{\tabcolsep}{3pt}
\renewcommand{\arraystretch}{1.08}
\begin{tabular}{|l|c|c|c|c|}
\hline
\textbf{Planner} & \(\boldsymbol{\mathcal{W}_{\mathrm{eval}}}\) &
\(\boldsymbol{P_{\mathrm{out}}(0.40)\downarrow}\) &
\(\boldsymbol{R_{\mathrm{gain}}\uparrow}\) &
\(\boldsymbol{R_{\mathrm{sw}}\downarrow}\) \\
\hline
Fixed narrow & \(\{1\}\) & 0.0070 & 0.9631 & 0.2346 \\
\hline
Fixed medium & \(\{3\}\) & 0.0051 & 0.9829 & 0.1862 \\
\hline
Fixed wide & \(\{5\}\) & 0.0043 & 0.9895 & 0.1234 \\
\hline
Adaptive BeamGuard & \(\{1,3,5\}\) & 0.0060 & 0.8948 & 0.1338 \\
\hline
\end{tabular}

\vspace{1pt}
\begin{minipage}{0.95\columnwidth}
\footnotesize
\textit{Note:} All rows use the same trained forecaster, test split, candidate-center construction, and forecast outputs; only the admissible virtual-beamwidth set differs. \(P_{\mathrm{out}}(0.40)\) denotes the threshold-based empirical outage probability, \(R_{\mathrm{gain}}\) the normalized covered-power gain ratio, and \(R_{\mathrm{sw}}\) the beam-center switching rate. Virtual beamwidth denotes adjacent-beam coverage within the finite ordered codebook and does not imply physical analog beam-pattern widening.
\end{minipage}
\end{table}
The fixed-width ablation shows that virtual beamwidth, defined as adjacent-beam coverage within the finite ordered codebook, materially affects reliability and control continuity. Under the evaluated setting, wider fixed coverage reduces both threshold-based empirical outage and beam-center switching. Adaptive BeamGuard, by contrast, jointly selects the beam center and virtual beamwidth using the predicted posterior, auxiliary power-based gain, risk penalty, beamwidth cost, and switching cost. The planner therefore provides a configurable mechanism for navigating the reliability--gain--switching tradeoff rather than committing to a single fixed coverage choice. This ablation reinforces the interpretation of BeamGuard as a forecasting-and-control framework and motivates future work on physical wide-beam synthesis and hardware-aware gain modeling.

\begin{table*}[!t]
\caption{Predictor Baselines Under Reduced and Full In-Band Beam Budgets}
\label{tab:baselines}
\centering
\scriptsize
\setlength{\tabcolsep}{2.4pt}
\renewcommand{\arraystretch}{1.08}
\begin{tabular}{|p{84pt}|p{58pt}|p{18pt}|p{32pt}|p{32pt}|p{32pt}|p{34pt}|p{34pt}|}
\hline
\textbf{Model} & \textbf{Input} & \(\boldsymbol{L}\) &
\textbf{Top-1} & \textbf{Top-3} & \textbf{Top-5} & \textbf{NLL} & \textbf{ECE} \\
\hline
Persistence & Beam history & -- & 0.268 & 0.579 & 0.745 & 2.954 & 0.177 \\
\hline
Sensor CNN & C+R+Lid+G & 0 & 0.262 & 0.578 & 0.761 & 2.565 & 0.087 \\
\hline
Vision-position proxy & C+G & 0 & 0.379 & 0.739 & 0.877 & 1.943 & 0.053 \\
\hline
LiDAR-position proxy & Lid+G & 0 & 0.357 & 0.712 & 0.864 & 2.014 & 0.049 \\
\hline
Pwr MLP & Pwr & 16 & 0.244 & 0.521 & 0.674 & 2.656 & 0.096 \\
\hline
Pwr MLP & Pwr & 64 & 0.278 & 0.585 & 0.737 & 2.371 & 0.060 \\
\hline
G+Pwr GRU & G+Pwr & 16 & 0.312 & 0.635 & 0.789 & 2.301 & 0.056 \\
\hline
G+Pwr GRU & G+Pwr & 64 & 0.307 & 0.641 & 0.790 & 2.283 & 0.062 \\
\hline
G+Pwr LSTM & G+Pwr & 16 & 0.313 & 0.637 & 0.794 & 2.304 & 0.058 \\
\hline
G+Pwr LSTM & G+Pwr & 64 & 0.328 & 0.669 & 0.811 & 2.189 & 0.056 \\
\hline
G+Pwr CNN & G+Pwr & 64 & 0.373 & 0.752 & 0.880 & 1.894 & 0.050 \\
\hline
BeamGuard G+Pwr & G+Pwr & 64 &
\textbf{0.405} & \textbf{0.791} & \textbf{0.910} & \textbf{1.784} & \textbf{0.046} \\
\hline
\end{tabular}

\vspace{2pt}
\begin{minipage}{0.97\textwidth}
\footnotesize
\textit{Note:} Results are reported on the joint day--night benchmark. For \(L<64\), only \(L\) entries of the 64-beam mmWave power vector are observed. C+G and Lid+G are same-protocol sensing-aided proxy baselines representing vision-position and LiDAR-position beam-prediction settings, respectively. Bold values indicate the best performance within the matched G+Pwr, \(L=64\) predictor comparison; lower NLL and ECE are better. The full-hybrid complete-system anchor is reported separately in Table~\ref{tab:anchor}.
\end{minipage}
\end{table*}
Table~\ref{tab:baselines} compares controlled same-protocol predictor baselines under the joint day--night benchmark. The added C+G and Lid+G rows provide numerical references for vision-position and LiDAR-position sensing-aided beam-prediction settings under the same BeamGuard split, horizon, codebook, calibration, and metric protocol. The BeamGuard G+Pwr row remains bolded because it is the strongest model within the matched G+Pwr, \(L=64\) predictor comparison. The full-hybrid complete-system anchor is reported separately in Table~\ref{tab:anchor}, since it evaluates the complete multimodal configuration together with planner-level reliability and latency metrics.

The fixed-width experiment isolates the effect of adjacent-beam coverage, while Table~\ref{tab:planner_controller_ablation} evaluates alternative controller rules using identical forecaster outputs. All variants use the same checkpoint, selected beam posterior, auxiliary power prediction, and test windows. The comparison therefore separates controller-side behavior from predictor performance and contrasts greedy fixed-width selection, a widest-safe heuristic, adaptive control without the explicit risk term, and the complete risk-aware planner.

\begin{table*}[!t]
\caption{Same-Forecaster Greedy and Risk-Ablation Controller Comparison}
\label{tab:planner_controller_ablation}
\centering
\footnotesize
\setlength{\tabcolsep}{3pt}
\renewcommand{\arraystretch}{1.08}
\begin{tabular}{|l|l|c|c|c|}
\hline
\textbf{Controller} &
\textbf{Controller Rule} &
\(\boldsymbol{P_{\mathrm{out}}(0.40)\downarrow}\) &
\(\boldsymbol{R_{\mathrm{gain}}\uparrow}\) &
\(\boldsymbol{R_{\mathrm{sw}}\downarrow}\) \\
\hline
Greedy Top-1 &
Top-1 center, fixed \(w=1\) &
0.0070 & 0.9631 & 0.2346 \\
\hline
Greedy Top-1 &
Top-1 center, fixed \(w=3\) &
\textbf{0.0057} & 0.9325 & 0.2346 \\
\hline
Widest-safe heuristic &
Widest risk-feasible action &
0.0064 & 0.8886 & 0.2346 \\
\hline
No-risk greedy &
Adaptive action, risk term removed &
0.0060 & 0.9420 & 0.1545 \\
\hline
Risk-aware BeamGuard &
Complete risk-aware planner &
0.0060 & 0.9056 & \textbf{0.1332} \\
\hline
\end{tabular}

\vspace{2pt}
\begin{minipage}{0.97\textwidth}
\footnotesize
\textit{Note:} All controller variants use the same trained forecaster, checkpoint, \(940\) test windows, selected beam posterior, and auxiliary power prediction. The common predictor Top-1 accuracy is \(0.3989\); only the controller decision rule changes. Bold entries identify the best value in each reported metric.
\end{minipage}
\end{table*}
Table~\ref{tab:planner_controller_ablation} confirms that controller logic affects communication-level behavior independently of predictor accuracy. Greedy Top-1 selection with \(w=1\) retains the highest covered-power gain but incurs the highest outage and switching rate, while fixed \(w=3\) achieves the lowest mean outage without reducing beam-center switching. Removing the explicit risk term gives the same mean outage as the complete BeamGuard planner but produces a higher switching rate. The risk-aware planner reduces switching by \(13.8\%\) relative to the no-risk controller and by approximately \(43.2\%\) relative to the greedy fixed-width controllers, although this improvement is accompanied by a lower covered-power gain ratio. The planner's contribution under the evaluated setting is therefore a more stable and explicitly configurable reliability--gain--switching tradeoff rather than uniform dominance in every individual metric.
\vspace{-2mm}
\subsection{Predictor Baselines}
\label{subsec:predbaseline}

Table~\ref{tab:baselines} presents controlled same-protocol predictor comparisons under the joint day--night benchmark. The input modality set and beam-observation budget \(L\) are reported explicitly so that architectural performance is not conflated with measurement availability. The added C+G and Lid+G rows provide numerical references for vision-position and LiDAR-position sensing-aided beam-prediction settings, respectively. These proxy baselines are evaluated using the same data splits, history length \(H=8\), prediction horizon \(T=5\), \(M=64\) beam target, calibration procedure, and evaluation metrics as BeamGuard.

The BeamGuard G+Pwr row is bolded because it provides the strongest performance within the matched G+Pwr, \(L=64\) predictor comparison, where all competing models use the same GPS and full in-band power inputs. This controlled comparison isolates the contribution of BeamGuard's temporal forecasting architecture from differences in modality availability or beam-observation budget. The strong ranked prediction of G+Pwr is consistent with the direct relationship between the measured mmWave power response, vehicle position, and the future best-beam index.

The full-hybrid result in Table~\ref{tab:anchor} is reported separately as the complete-system reference used in the abstract. It evaluates the full C+R+Lid+G+Pwr configuration together with planner-level outage, gain, switching, and latency metrics, whereas Table~\ref{tab:baselines} evaluates predictor performance under controlled input settings. Adding camera, radar, and LiDAR does not necessarily maximize Top-\(K_{\mathrm{acc}}\) accuracy because these high-dimensional sensing streams can introduce domain-dependent scene variation and are less directly coupled to the exact beam index than the in-band power response. Their additional environmental context nevertheless remains useful for reliability-oriented beam management when forecast uncertainty is converted into adaptive virtual beamwidth actions. The two tables therefore distinguish matched-input prediction strength from complete-system forecasting-and-control performance without implying that full multimodal fusion uniformly maximizes every predictor metric.
\begin{table}[!t]
\caption{Runtime and Complexity of the BeamGuard Hybrid Anchor}
\label{tab:runtime}
\centering
\footnotesize
\setlength{\tabcolsep}{3pt}
\renewcommand{\arraystretch}{1.08}
\begin{tabular}{|p{115pt}|p{95pt}|}
\hline
\textbf{Metric} & \textbf{Value} \\
\hline
Profiling platform &
Apple M2 Pro, 16-GB unified memory, PyTorch MPS \\
\hline
Total parameters & 39.56M \\
\hline
Checkpoint size & 151.17 MB \\
\hline
Inference pooling batch size & 1 \\
\hline
Predictor latency & 76.37 ms \\
\hline
Planner latency & 1.49 ms \\
\hline
End-to-end latency & 77.86 ms \\
\hline
Throughput & 12.84 windows/s \\
\hline
\end{tabular}
\end{table}

\subsection{Runtime, Practical Feasibility, and Key Findings}

Table~\ref{tab:runtime} summarizes the model complexity and single-window inference runtime of the full-hybrid anchor. Runtime profiling uses a batch size of \(1\) on an Apple M2 Pro platform with \(16\)~GB unified memory and PyTorch MPS acceleration. The model contains approximately \(39.56\) million parameters and has a checkpoint size of \(151.17\)~MB. The predictor requires \(76.37\)~ms per window, while the planner adds \(1.49\)~ms per window, corresponding to approximately \(1.9\%\) of the total measured latency. The resulting end-to-end latency is \(77.86\)~ms per window, with a throughput of \(12.84\) windows/s. Since this latency is below the \(100\)-ms experimental decision interval, the implementation is compatible with the receding-horizon cadence evaluated on the reported Apple MPS platform. These measurements are platform-specific and do not constitute a hardware-independent real-time deployment guarantee.

The reported runtime does not directly characterize an embedded roadside unit or a Jetson-class accelerator. Embedded inference latency depends on the accelerator generation, power mode, numerical precision, inference runtime, memory bandwidth, and whether sensor preprocessing is executed on the same device. The full-hybrid configuration is the most computationally demanding operating point because it includes the camera, radar, and LiDAR spatial encoders in addition to GPS, power processing, multimodal fusion, and temporal forecasting. The measured latency breakdown shows that the predictor path accounts for approximately \(98.1\%\) of the end-to-end latency, whereas the risk-aware planner contributes only about \(1.9\%\).

The evaluated operating regimes provide an accuracy--complexity tradeoff that is relevant to embedded deployment. G+Pwr removes the high-dimensional camera, radar, and LiDAR encoders and achieves the strongest ranked prediction within the matched G+Pwr comparison, making it an attractive low-complexity operating point when communication-side power measurements are available. G+Lid+Pwr adds geometric context while retaining fewer sensing branches than the full-hybrid model. The full C+R+Lid+G+Pwr configuration provides the richest environmental context and can improve reliability under more difficult held-out conditions, but it also has the highest encoder and memory demand. Because these variants have not yet been profiled on an embedded accelerator, this comparison is interpreted as an architecture- and accuracy-based deployment tradeoff rather than as a measured Jetson latency comparison.

The overhead-aware formulation in this work primarily concerns communication overhead. The observation budget \(L\) controls how many in-band beam-power measurements are required, while sensor-only and partial-observation modes reduce or avoid exhaustive beam probing. This communication-overhead reduction is distinct from computational overhead on the roadside processor. Establishing the latter requires hardware-specific measurements of latency, memory, energy consumption, and thermal behavior. A deployment-oriented extension will therefore evaluate the full-hybrid and lightweight operating modes on embedded accelerators using batch-one inference, mixed-precision and quantized execution, lightweight sensing backbones, pruning or distillation, and hardware-optimized runtimes. The resulting evaluation will quantify median and tail latency, peak memory usage, power consumption or energy per inference, together with the corresponding changes in Top-\(K_{\mathrm{acc}}\) accuracy, NLL, ECE, threshold-based empirical outage, gain ratio, and beam-center switching rate.

Across the evaluated software and communication settings, BeamGuard is assessed as a beam-management framework rather than as an isolated beam classifier. Ranked prediction accuracy is informative, but threshold-based outage, covered-power gain, and switching behavior determine whether the predicted posterior produces useful control actions. The beam-budget and sampling-policy results show that partial in-band observations can retain much of the full-observation benefit under more than one mask-generation rule. The held-out S31/S34 tests expose the difficulty of broader scenario transfer, while the fixed-width and same-forecaster controller ablations clarify the separate effects of adjacent-beam coverage and controller logic. Finally, the risk-budget sweep shows that the principal effect of varying \(\beta\) over the tested range is improved beam-center continuity with only marginal changes in outage and gain. These findings support the value of coupling multimodal forecasting with explicit virtual beamwidth control under the evaluated protocols, while embedded hardware validation remains necessary before deployment-level latency and energy claims can be made.
\vspace{-2mm}
\subsection{Limitations and Future Work}

The present study evaluates BeamGuard at the codebook-control level using virtual beamwidths defined by adjacent-beam coverage, rather than physically synthesized analog wide beams. The primary development and evaluation protocols use Scenarios 32 and 33, while Scenarios 31 and 34 serve only as compact held-out stress tests; the pronounced degradation on these scenarios shows that the results do not establish universal cross-scenario generalization. The experiments also consider a single V2I link, predefined uniform and local probing policies, and software profiling on Apple MPS rather than embedded roadside hardware. Future work will extend the framework to physical wide-beam synthesis, multi-user and multi-RSU coordination, online domain adaptation, and hardware-aware deployment on roadside edge platforms.
\vspace{-2mm}
\section{Conclusion}

BeamGuard advances reliable and overhead-aware beam management for 6G mmWave V2I links by coupling multimodal future beam forecasting with risk-aware adaptive virtual beamwidth control. The framework converts camera, radar, LiDAR, GPS, and optional mmWave power observations into beam actions that balance predicted risk, beamforming gain, beamwidth, and switching cost. It supports sensor-only, partial in-band, and full hybrid operation through explicit beam-observation masking. Experiments on DeepSense 6G Scenarios 32 and 33 demonstrate strong ranked beam forecasting, low threshold-based outage, and practical tradeoffs between beam-training overhead and link reliability. Sampling-policy, fixed-width, same-forecaster controller, and risk-budget ablations separate the contributions of observation pattern, adjacent-beam coverage, and controller logic. Held-out tests on Scenarios 31 and 34 further reveal substantial scenario shift, underscoring the need for domain-robust learning and calibration. Overall, BeamGuard provides a technically explicit forecasting-and-control framework for multimodal 6G V2I beam management while identifying physical beam synthesis, broader scenario validation, multi-user coordination, and embedded deployment as important next steps.
\vspace{-2mm}

\vfill

\end{document}